\documentclass[letterpaper, 10 pt, conference]{ieeeconf}  

\IEEEoverridecommandlockouts                              

\usepackage{lineno,hyperref}
\modulolinenumbers[5]

\usepackage{graphicx} 
\usepackage{epsfig} 
\usepackage{amsmath} 
\usepackage{amssymb}  
\usepackage{epstopdf}
\usepackage{mathtools}
\usepackage{multirow}
\usepackage{array}
\usepackage{siunitx}
\usepackage{tikz}
\usepackage{mathrsfs}
\usepackage{tabularx}
\usepackage{url}
\usetikzlibrary{calc}
\usetikzlibrary{positioning}
\usetikzlibrary{arrows}
\usetikzlibrary{shapes}
\usetikzlibrary{fit}

\usepackage{algpseudocode}
\usepackage{float}
\usepackage{xcolor}
\usepackage{microtype}
\usepackage{booktabs}

\usepackage{svg}
\usepackage{psfrag}

\usepackage{pgfplots}
\usetikzlibrary{calc}
\pgfplotsset{compat=1.8}

\definecolor{line1}{cmyk}{0.02,0.16,1,0}
\definecolor{line2}{cmyk}{0.86,1,0.03,0.01}

\graphicspath{{Figures/}}

\newcommand{\nth}[1]{$#1^{\textrm{th}}$}
\newcommand{\actseq}[1]{\boldsymbol{\gamma_{t}[#1]} = \left\{\gamma_{t}[#1],\,\gamma_{t+1}[#1],\,\cdots,\,\gamma_{t+N-1}[#1]\right\}}
\newcommand{\actseqk}[2]{\boldsymbol{\gamma^{\left(#2\right)}_{t}[#1]} = \left\{\gamma^{\left(#2\right)}_{t}[#1],\,\gamma^{\left(#2\right)}_{t+1}[#1],\,\cdots,\,\gamma^{\left(#2\right)}_{t+N-1}[#1]\right\}}
\newcommand{\actk}[3]{\gamma^{\left(#2\right)}_{#3}[#1]}
\newcommand{\actt}[3]{\gamma^{}[#1](#3)}
\newcommand{\actseqkt}[3]{\left\{\gamma^{\left(#2\right)}_{t}[#1],\,\gamma^{\left(#2\right)}_{t+1}[#1],\,\cdots,\,\gamma^{\left(#2\right)}_{t+#3}[#1]\right\}}

\newcommand{\optactseq}[1]{\boldsymbol{\gamma^{*}_{t}[#1]} = \left\{\gamma^{*}_{t}[#1],\,\gamma^{*}_{t+1}[#1],\,\cdots,\,\gamma^{*}_{t+N-1}[#1]\right\}}
\newcommand{\optact}[1]{\gamma^{*}_{t}[#1]}

\newcommand{\actseqktnopar}[3]{\gamma^{\left(#2\right)}_{t}[#1],\,\gamma^{\left(#2\right)}_{t+1}[#1],\,\cdots,\,\gamma^{\left(#2\right)}_{t+#3}[#1]}

\begin{document}
	
	\title{Adaptive Robust Game-Theoretic Decision Making for Autonomous Vehicles}

	\thispagestyle{empty}

	\author{
		Gokul S.~Sankar and  Kyoungseok Han 
		\thanks{Gokul~S.~Sankar and  Kyoungseok Han are with the Department of Aerospace Engineering, University of Michigan, Ann Arbor, MI, USA. Emails: {\tt\small ggowrisa@umich.edu} (G.~Sankar) and  
			{\tt\small kyoungsh@umich.edu} (K.~Han) }%
	}

	\maketitle

	\begin{abstract}
		
		In a typical traffic scenario, autonomous vehicles are required to share the road with other road participants, e.g., human driven vehicles, pedestrians, etc. To successfully navigate the traffic, an adaptive robust level-\textit{k} framework can be used by the autonomous agents to categorize the agents based on their depth of strategic thought and act accordingly. However, mismatch between the vehicle dynamics and its predictions, and improper classification of the agents can lead to undesirable behavior or collision. Robust approaches can handle the  uncertainties, however, might result in a conservative behavior of the autonomous vehicle. This paper proposes an adaptive robust decision making strategy for autonomous vehicles to handle model mismatches in the prediction model while utilizing the confidence of the driver behavior to obtain less conservative actions. The effectiveness of the proposed approach is validated for a lane change maneuver in a highway driving scenario.
	
	\end{abstract}

	
	\section{Introduction}
	\label{sec:intro}

	Connected and automated vehicles (CAVs) is a disruptive transportation technology that has the potential to change our habits and provide great safety benefits. Despite many recent advances in CAV technologies, full automation of the vehicles that provides better or at least as good driving proficiency as compared to the human drivers are still flawed to be deployed in the market \cite{Okuda2014}. One of the most significant challenges is to plan the motion in mixed traffic scenarios, where the autonomous vehicles (AVs) coexist with the human driven vehicles (HVs), bike riders and pedestrians \cite{Lazar2018}. In particular, describing the human decision making process is difficult since humans do not always exhibit optimized behavior due to limited rationality \cite{Griffiths2015}.

	The interactions between agents, either vehicle-to-vehicle or vehicle-to-pedestrian, in a mixed traffic scenario, are accounted by considering the stochastic reachable sets of the vehicles \cite{Althoff2009}. The desired path of each agent is assumed to be known, however, might be imperfect. Alternatively, game-theoretic framework can be used to m ake strategic decisions that handle interactions in the multi agent mixed traffic scenario. 

	In the level-k game-theoretic approach, the agents are categorized into hierarchical structure of their cognitive abilities \cite{Stahl1993}. The interactions are modeled taking into account the rationality of the other agents. This might lead to a less conservative action the reachable set method. Previously, the game-theoretic modeling approach has been applied to highway driving scenario in \cite{Li2017}, and to an unsignalized intersection in \cite{Li2018, Tian2018}. A multi-model strategy is used that assigns an agent to multiple levels in the hierarchical structure  with certain confidence from the perspective of the AVs. The estimate of the driver level and its probability of being at a certain level is updated at every time step by the autonomous agents.
	
	Although these approaches effectively describe rational decision making for the HVs, uncertainties due to the simple vehicle dynamic model used within the framework, and an improper estimation of the driver level of the other agents, are not considered. These uncertainties in the position of the HVs in the decision making process might lead to collision. A high fidelity dynamic model can be used to eliminate certain level of uncertainty but is accompanied by increased computational burden. Also, most likely, the interactions between agents in a given traffic scenario might be short for the AVs to estimate an accurate driver model. In \cite{Althoff2009}, the uncertainty in the interactions between the agents are described by probabilistic deviations from the desired path, however, as mentioned earlier, the desired path assumed might be inaccurate.

	Robust model-based approaches namely, feedback min-max model predictive control (MPC) \cite{Scokaert1998}, tube MPC \cite{Mayne2005} and constraint-tightening methodology \cite{Richards2003} have been suggested and well established for the last couple of decades. These robust approaches provide constraint satisfaction guarantees for the uncertainties originating from a bounded and compact disturbance set. The constraint tightening approach has no additional online computational load and been successfully implemented in real time automotive applications \cite{Sankar2019, Sankar2019a} and aerospace applications \cite{Richards2003}. However, due to its simplicity in handling the discrete input actions considered in this work, min-max strategy is employed to provide robustness to the uncertainties described above. 

	The min-max strategy considers the worst-case disturbance affecting the behavior/performance of the system and provides control actions to mitigate the effect of the worst-case disturbance. Therefore, these control actions might lead to a conservative behavior if the disturbance set size that should be handled is large. In \cite{Jin2019}, a conditional value-at-risk objective function was used to account for the model mismatch. A constant confidence level is introduced, which describes the degree of aggressiveness of the HV. However, a \textit{priori} assumption on the confidence level of the human drivers may not be realistic. To provide robustness against all possible traffic situations, a conservative driving policy has been suggested for the AV by \cite{Claussmann2015, Brechtel2014}, However, conservative actions can have adverse effects on the traffic, for instance, lead to a road congestion, disharmony with the other road participants. Less conservative motion can be planned for the AV  when the behavior of human vehicles are predicted \cite{Sadigh2016}.

	In this paper, an adaptive robust approach has been proposed to provide less conservative actions to the AVs to perform a desired maneuver in the multi agent traffic scenario. The min-max approach anticipates the uncertainties originating from a disturbance set while computing the control action. The adaptive strategy proposed in this paper modifies the size of the disturbance sets of the other interacting vehicles in accordance to the belief of the aggressiveness of the corresponding human drivers. The proposed approach can provide `balanced' control actions for the AV which is more conservative than the nominal (non-robust) strategy to handle model mismatch and on the other hand, less conservative than traditional robust approach by exploiting the confidence on the estimated driver model. The proposed approach is validated through simulation studies for performing a lane changing maneuver on a highway driving scenario with multiple agents.
	

	\subsection{Notation}
	The symbol $\mathbb{Z}_{\left[a,\, b\right]}$ denotes a set of consecutive integers from $a$ to $b$ and $2\mathbb{Z}^{+}$ denotes set of positive even integers; $\emptyset$ denotes an empty set. For a vector $x$, $x>0$ denotes element-wise inequality. The operator $\oplus$ denotes the Minkowski addition, defined for the sets $\mathcal{A}$ and $\mathcal{B}$ as $	\mathcal{A}\oplus\mathcal{B}\coloneqq\left\{ a+b | a\in\mathcal{A}\,\forall b\in\mathcal{B}\right\}$.

	\begin{figure}
		\begin{centering}
			\begin{tikzpicture}
			
			\node (rect) [draw =none, very thin, fill = white, minimum width = 8.25cm, minimum height = 3cm, inner sep = 0pt] {};
			
			\node (off_road_north) [draw, very thick, fill = black, minimum width = 8.25cm, minimum height = 0.1mm, inner sep = 0pt, above  = 0mm of rect] {};
			
			\node (off_road_south) [draw, very thick, fill = black!60!green, minimum width = 8.25cm, minimum height = 0.1mm, inner sep = 0pt, below  = 0mm of rect] {};

			\draw  [loosely dashed, thick] (rect.west) -- (rect.east);
			\node (blue)	[below left = -1.25cm and -2.2cm of rect]{\includegraphics[scale=0.1]{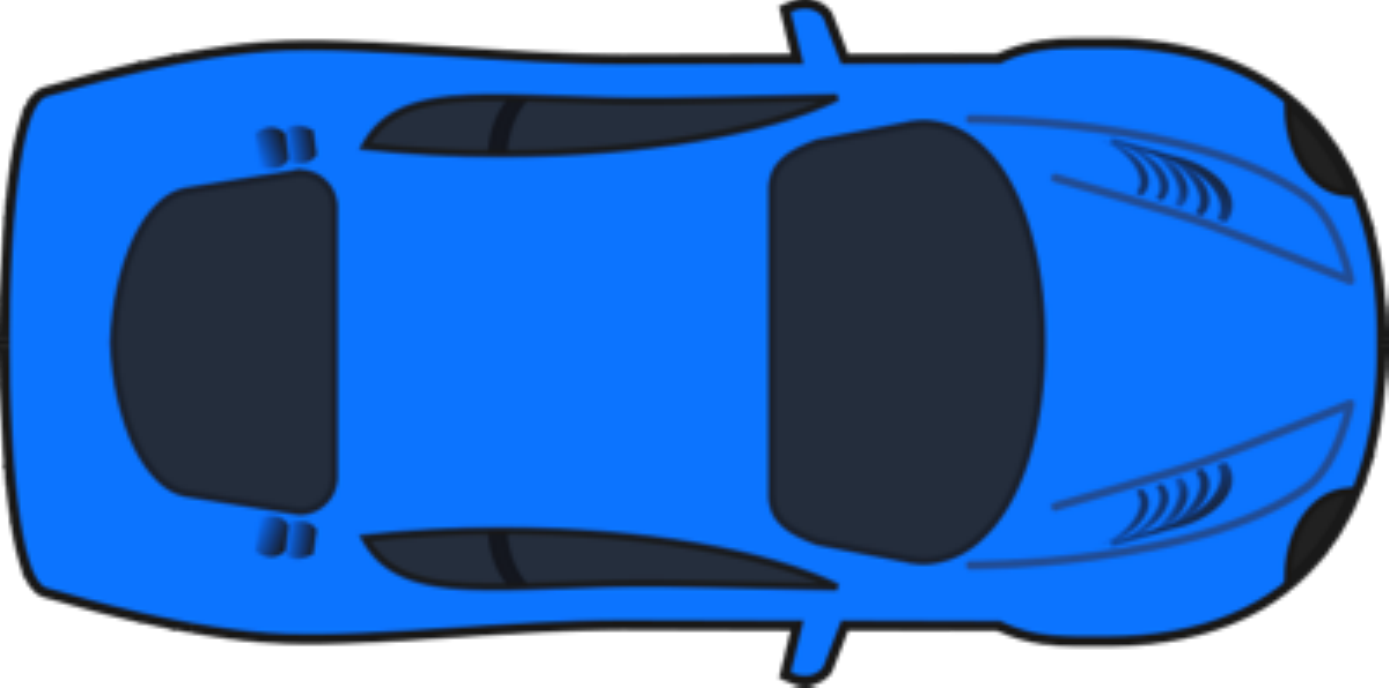}};
			\node (red1) [above left = -1.25cm and -3.2cm of rect] {\includegraphics[scale=0.1]{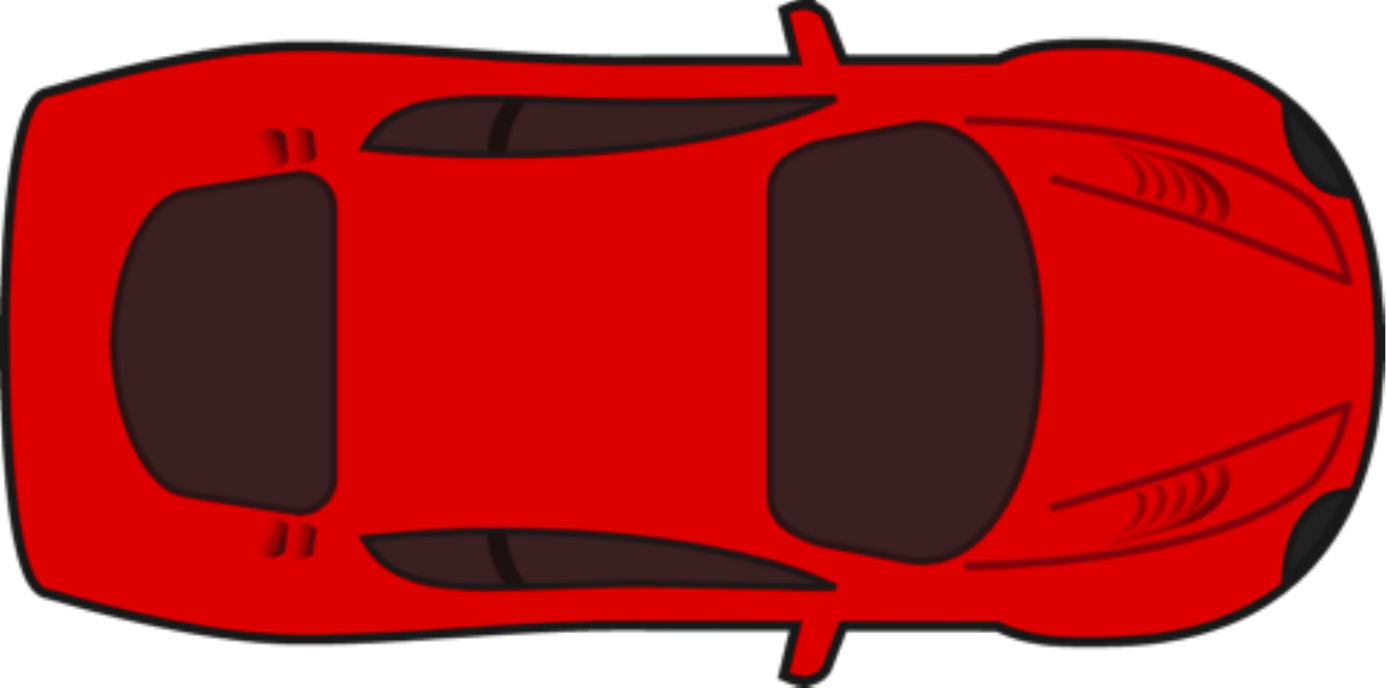}};
			\node (red2)	[above right = -2.75cm and -2cm of rect]{\includegraphics[scale=0.1]{red_car.pdf}};
			
			\node (mid3)	[above left = -1.58cm and -5.9cm of rect, minimum width = 0cm, minimum height = 0cm, inner sep = 0pt] {};
			\node (mid2)	[above left = -1.56cm and -4.9cm of rect, minimum width = 0.cm, minimum height = 0.cm, inner sep = 0pt] {};
			\node (mid1)	[above left = -1.56cm and -3.9cm of rect, minimum width = 0.cm, minimum height = 0.cm, inner sep = 0pt] {};

			\node (off_road_north) [draw = none, very thick, fill = black!60!green, minimum width = 8.25cm, minimum height = 1cm, inner sep = 0pt, above  = 0mm of rect, fill opacity=0.6] {};
			
			\node (off_road_south) [draw = none, very thick, fill = black!60!green, minimum width = 8.25cm, minimum height = 1cm, inner sep = 0pt, below  = 0mm of rect, fill opacity=0.6] {};

			\node (final)	[draw, circle, above right = -0.83cm and -0.5cm of rect, minimum width = 0.cm, minimum height = 0.cm, inner sep = 0pt] {};
			
			\node (final3)	[draw,  above right = -0.83cm and -1.5cm of rect, minimum width = 0.cm, minimum height = 0.cm, inner sep = 0pt] {};
			\node (final2)	[draw, circle, above right = -0.83cm and -2.5cm of rect, minimum width = 0.cm, minimum height = 0.cm, inner sep = 0pt] {};
			\node (final1)	[draw, circle, above right = -0.83cm and -3.5cm of rect, minimum width = 0.cm, minimum height = 0.cm, inner sep = 0pt] {};

			\node (breakaway1)	[draw, circle,  below left = -0.83cm and -3cm of rect, minimum width = 0.cm, minimum height = 0cm, inner sep = 0pt]{};
			\node (breakaway2)	[draw, circle,  below left = -0.83cm and -4cm of rect, minimum width = 0.cm, minimum height = 0cm, inner sep = 0pt]{};
			\node (breakaway3)	[draw, circle,  below left = -0.83cm and -5cm of rect, minimum width = 0.cm, minimum height = 0cm, inner sep = 0pt]{};		
			
			\draw [rounded corners,color=cyan,line width=3pt] (blue) .. controls (breakaway1) ..  (mid1) .. controls (final1) .. (final2) -- (final);
			\draw [rounded corners,color=line1,line width=3pt] (blue) .. controls (breakaway2) ..  (mid2) .. controls (final2) .. (final3) -- (final);
			\draw [rounded corners,color=line2,line width=3pt] (blue) .. controls (breakaway3) ..  (mid3) .. controls (final3) .. (final) ;

			\node [draw, fill = none, minimum width = 1.5cm, minimum height = 0.75cm, inner sep = 0pt, below left = -2.32cm and -2.58cm of blue, thin, dash dot, color = red] {};
		
			\node [draw, fill = none, minimum width = 1.5cm, minimum height = 0.75cm, inner sep = 0pt, below left = -0.86cm and -1.57cm of red2, thin, dash dot, color = red] {};
			
			\node [draw, fill = none, minimum width = 1.5cm, minimum height = 0.75cm, inner sep = 0pt, below left = -0.86cm and -1.57cm of blue, thin, dash dot, color = blue] {};
			
			\node[above left = -1.15cm and -2cm of blue,]{A};
			\node[above left = -1.15cm and -2.75cm of blue,]{A1};
			\node[above left = -1.15cm and -3.75cm of blue,]{A2};
			\node[above left = -1.15cm and -4.75cm of blue,]{A3};
			\node[above left = 1.15cm and -4.5cm of blue,]{B1};
			\node[above left = 1.15cm and -5.5cm of blue,]{B2};
			\node[above left = 1.15cm and -6.5cm of blue,]{B3};
			\node[above left = 1.15cm and -7.5cm of blue,]{B};

			\node(car1) [below left = -0.67cm and 0.1cm of red2, inner sep = 0pt, draw, circle, minimum height = 0.5cm, minimum width = 0.5cm] {1};
			\node(car2) [below left = -0.67cm and 0.1cm of blue, inner sep = 0pt, draw, circle, minimum height = 0.5cm, minimum width = 0.5cm] {2};
			\node(car3) [below left = -0.67cm and 0.1cm of red1, inner sep = 0pt, draw, circle, minimum height = 0.5cm, minimum width = 0.5cm] {3};
			
			\end{tikzpicture}
			\par
		\end{centering}
	\protect\caption{Sample lane changing sequences that can be chosen by the autonomous vehicle (blue) based on the motion of other non-autonomous vehicles (red) are $\left\{\left(A,A1\right),\,\left(A1,B1\right),\,\left(B1,B\right)\right\}$, $\left\{\left(A,A2\right),\,\left(A2,B2\right),\,\left(B2,B\right)\right\}$ and $\left\{\left(A,A3\right),\,\left(A3,B3\right),\,\left(B3,B\right)\right\}$.}
	\label{fig:lane_changing}
	\end{figure}


	\section{Problem statement}
	\label{sec:problem_statement}

	Consider a highway driving traffic scenario shown in the Fig.~\ref{fig:lane_changing} with human driven vehicles shown in red and the autonomous vehicle in blue. The goal of the AV is to perform a lane change. A subset of the possible lane changing trajectories are shown in the Fig.~\ref{fig:lane_changing}. The trajectory $\left\{\left(A,A1\right),\,\left(A1,B1\right),\,\left(B1,B\right)\right\}$ is considered to be an aggressive maneuver as the lane change occurs closer to the human vehicle $3$ and might lead to collision. On the other hand, $\left\{\left(A,A3\right),\,\left(A3,B3\right),\,\left(B3,B\right)\right\}$ is considered conservative as the AV completes the maneuver by performing the lane change farthest from the HV 3. Being conservative can lead to road congestion and disharmony as mentioned earlier. Therefore, it is desirable to develop an adaptive control strategy for the AV that is able to perform the desired maneuver with reduced conservatism whilst planning a safe motion to avoid collision as in $\left\{\left(A,A2\right),\,\left(A2,B2\right),\,\left(B2,B\right)\right\}$. The strategy should be able to  modify the behavior of the AV according to the confidence on the estimated behavior of the interacting vehicles.


	\section{Vehicle dynamic model}
	\label{sec:model}

	The  vehicle dynamics are represented by the following discrete model \cite{Kong2015}
	\begin{subequations}
		\begin{align}
			x\left(t+1\right) & = 	x\left(t\right) + v\left(t\right) \cos\left(\psi \left(t\right) + \beta \left(t\right) \right) \Delta t + w_x\left(t\right)\\ 
			y\left(t+1\right) & = 	y\left(t\right) + v\left(t\right) \sin\left(\psi \left(t\right) + \beta \left(t\right) \right) \Delta t + w_y\left(t\right) \\ 
			\psi\left(t+1\right) & = 	\psi\left(t\right) + \frac{v\left(t\right)}{l_r} \sin\left( \beta \left(t\right) \right) \Delta t \\ 
			v\left(t+1\right) & = 	v\left(t\right) + a\left(t\right)  \Delta t\\
			\beta\left(t\right) & = \arctan\left(\frac{l_r}{l_r+l_f}\tan\left(\delta_f\left(t\right)\right)\right),
		\end{align}
	\label{eq:mod}
	\end{subequations}

	\noindent where  $t$ denotes the discrete time instant; the pair $\left(x\left(t\right),\,y\left(t\right)\right)$  represent the global position of the center of mass of the vehicle; the vehicle's speed is denoted by $v\left(t\right)$; 	$\beta\left(t\right)$ is the angle of $v\left(t\right)$ with respect to the longitudinal axis of the vehicle; $\psi\left(t\right)$ denotes the vehicle’s yaw angle (the angle between the vehicle’s heading direction and the global x-direction); $a\left(t\right)$ denotes the vehicle’s acceleration at time $t$;  $\Delta t$ denotes the time step size; $\delta_f\left(t\right)$  represents the front steering angle; and  $l_f$  and $l_r$  are the distance of the center of the mass of the vehicle to the front and rear axles, respectively; $w_x\left(t\right)$  and $w_x\left(t\right)$ denote the uncertainty in the position of the center of mass, respectively. It is assumed that the uncertainties originate from a closed and compact set defined as 
	\begin{align}
		{\mathcal{W}} \coloneqq  \left\{ w = \left(w_x,\,w_y\right) |\zeta w\leq\theta,\,\zeta\in\mathbb{R}^{a\times 2},\,\theta\in\mathbb{R}^{b}\right\},
		\label{eq:dist_set}
	\end{align}
	where $a,\, b \in 2\mathbb{Z}^{+}$. The disturbance set is assumed to contain the origin. Furthermore, it is assumed that the rear wheels cannot be steered. Therefore, the control input to the model \eqref{eq:mod}, represented by $\gamma\left(t\right) = \left(a\left(t\right),\, \delta_f\left(t\right)\right)$, is the acceleration and front steering angle pair.


	\section{Robust game-theoretic decision making}
	\label{sec:controller}

	At each time instant, each vehicle selects an input pair from the finite action set, $ \boldsymbol{\Gamma} = \left\{\left(0,\, 0\right),\,\left(0,\, \delta_{f,\,\textrm{nom}}\right),\,\left(0,\, -\delta_{f,\,\textrm{nom}}\right),\,\left(a_{\textrm{nom}},\, 0\right),\,\left(-a_{\textrm{nom}},\, 0\right),\right.$ $\left.\left(a_{\max},\, 0\right),\,\left(-a_{\max},\, 0\right),\,\left(a_{\textrm{nom}},\, \delta_{f,\,\max}\right),\,\left(a_{\textrm{nom}},\, -\delta_{f,\,\max}\right)\right\}$, where $a_{\textrm{nom}}$, $\delta_{f,\,\textrm{nom}}$ and $a_{\max}$, $\delta_{f,\,\max}$ are the nominal and maximum acceleration, front steer angle, respectively. The inputs pairs in $ \boldsymbol{\Gamma}$ correspond to the actions, \{``maintain", ``turn slightly left", ``turn slightly right", ``accelerate", ``decelerate", ``maximum acceleration", ``maximum deceleration", ``turn left and accelerate", `` turn right and accelerate'' \}, respectively. The input pair to be applied at every time step is decided based on optimizing a reward function.

	\subsection{Reward function}
	\label{sec:reward_function}
	
	The decision making process of the vehicle in choosing the optimal input pair follows a receding horizon strategy. The nominal model used within the prediction horizon is given by
	\begin{subequations}
		\begin{align}
			x_{t+j+1} & = 	x_{t+j} + v_{t+j} \cos\left(\psi_{t+j} + \beta_{t+j} \right) \Delta t \\ 
			y_{t+j+1} & = 	y_{t+j} + v_{t+j} \sin\left(\psi_{t+j} + \beta_{t+j} \right) \Delta t  \\ 
			\psi_{t+j+1} & = 	\psi_{t+j} + \frac{v_{t+j}}{l_r} \sin\left( \beta _{t+j} \right) \Delta t \label{eq:psi_pred}\\ 
			v_{t+j+1} & = 	v_{t+j} + a_{t+j}  \Delta t \label{eq:v_pred}\\
			\beta_{t+j} & = \arctan\left(\frac{l_r}{l_r+l_f}\tan\left(\delta_{f,\,{t+j}}\right)\right) \label{eq:beta_pred},
		\end{align}
	\label{eq:nom_mod}
	\end{subequations}

	\noindent where $j \in \mathbb{Z}_{\left[0,\,N-1\right]}$  represents the prediction step, and $\gamma_{t+j} = \left(a_{t+j},\, \delta_{f,\,t+j}\right)$ denotes the input pair applied to \eqref{eq:nom_mod} at a prediction step $j$. A sequence of actions, $\boldsymbol{\gamma_{t}} = \left\{\gamma_{t},\,\gamma_{t+1},\,\cdots,\,\gamma_{t+N-1}\right\}$, is chosen that maximizes a cumulative reward given by
	\begin{align}
	\mathcal{R}\left(\boldsymbol{\gamma_{t}}\right) = \sum_{j=0}^{N-1} \lambda^{j} R_{t+j},
	\label{eq:cum_reward}
	\end{align}
	
	\noindent where $R_{t+j}$ is the stage reward at a prediction step $j$ determined at time step $t$ for an input, $\gamma_{t+j} \in  \boldsymbol{\Gamma}$; $\lambda \in \left[0,\,1\right]$ is the discount factor. By the receding horizon strategy, the input applied to \eqref{eq:mod}, $\gamma\left(t\right)$, is the first element of $\boldsymbol{\gamma_{t}}^* = \left\{\gamma_{t}^*,\,\gamma_{t+1}^*,\,\cdots,\,\gamma_{t+N-1}^*\right\}$ is applied at each time instant $t$. The stage reward at a prediction step $j$, $R_{t+j}$, is defined as
	 \begin{align}
	 R_{t+j} = \boldsymbol{{\alpha}}^T \boldsymbol{\phi_{t+j}},
	 \label{eq:stage_reward}
	 \end{align}
	
	\noindent where $\boldsymbol{{\phi}_{t+j}} = \left\{\phi_{1,\,{t+j}},\,\phi_{2,\,{t+j}},\,\cdots,\,\phi_{m,\,{t+j}}\right\}$ is the feature vector at step $j$ and the weights for these features are in  $\boldsymbol{{\alpha}} = \left\{\alpha_{1},\,\alpha_{2},\,\cdots,\,\alpha_{m}\right\}$, in which $\alpha_{i}>0,\,\forall \, i\in \mathbb{Z}_{\left[1,\,m\right]}$. For the lane changing scenario in  Fig.~\ref{fig:lane_changing}, the  features considered are described below. 
	
	Rectangular outer approximation of the geometric contour of each vehicle is considered as shown by the dash-dotted boxes in Fig.~\ref{fig:lane_changing}. This outer approximation is referred as the collision avoidance zone (c). The features, $\phi_{1,\,{t}},\,\phi_{2,\,{t}}$ and $\phi_{3,\,{t}}$, are indicator functions based on the collision avoidance zone of the vehicles that respectively characterize:
	\begin{itemize}
		\item Collision status - The intersection of the collision avoidance zone of the ego vehicle with that of any other vehicle indicates a collision or a danger of collision. If an overlap is detected then $\phi_{1,\,{t}}$ is assigned a value $-1$; and $0$, otherwise.
		
		\item On-road status - The intersection of the collision avoidance zone of the ego vehicle with that of green regions shown in Fig.~\ref{fig:lane_changing} indicates that the ego vehicle is outside the road boundaries. The feature $\phi_{2,\,{t}} = -1$ if an overlap is detected; $\phi_{2,\,{t}} = 0$, otherwise.
		
		\item Safe zone violation status - A safe zone (s) of a vehicle is a rectangular area that subsumes the collision avoidance zone of the vehicles with a safety margin. The safety margin is chosen based on the minimum distance to be maintained from the surrounding vehicles. If an overlap of the safe zone of the ego vehicle with that of another vehicle is detected then $\phi_{3,\,{t}}$ is assigned a value $-1$; and $0$, otherwise.

	\end{itemize}
	 
	The other features considered in this work characterize:
	\begin{itemize}
	 	\item Distance to objective - In order to encourage the ego vehicle to change lane and reach a reference point in the new lane, $\left(x^{\textrm{ref}},\,y^{\textrm{ref}}\right)$, the feature $\phi_{4,\,{t}} $ is defined as
	 	\begin{align}
	 		\phi_{4,\,{t}}  = -\left(\left|x_{t}-x^{\textrm{ref}}\right| + \left|y_{t}-y^{\textrm{ref}}\right|\right).
	 	\end{align}
	 	
	 	\item Distance to lane center - The feature, $\phi_{5,\,{t}} $, defined as 
	 	 \begin{align}
	 	 \phi_{5,\,{t}}  = - \left|y_{t}-y^{lc}\right|,
	 	 \end{align}
	 	 
	 	 \noindent where $y^{lc}$ is the y-coordinate of the center of the current lane that is included to encourage the ego vehicle to be at the middle of the current lane.
	 	 
	 	\item Velocity error - The deviation of the velocity of the ego vehicle from a reference velocity, $v^{\textrm{ref}}$, is described by the feature $\phi_{6,\,{t}}$ as
	 	\begin{align}
	 		\phi_{6,\,{t}} = - \left|v_{t}-v^{\textrm{ref}}\right|,
	 	\end{align}
	 	where the reference velocity is typically chosen as the legislated speed limit.
	\end{itemize}

	Intuitively, the first three features are prioritized to avoid actions taken by the controller leading to a catastrophe, and therefore, the tuning parameters are chosen as  
	\begin{equation}\label{eq:weights_order}
     \alpha_1, \alpha_2, \alpha_3 \gg \alpha_4, \alpha_5, \alpha_6. \nonumber
    \end{equation}

 	\subsection{Level-k framework}
	\label{sec:level_k}

	In a multi-agent traffic scenario, the interactive nature of the decision making process is taken into account by the features, $\phi_{1,\,t}$ and $\phi_{3,\,t}$, of the stage reward in \eqref{eq:stage_reward} that depend on the states of other vehicles. To compute the cumulative reward in \eqref{eq:cum_reward}, for a given sequence of actions of the \nth{l} autonomous vehicle, $\actseq{l}$, it is required to predict the actions of other agents, $\actseq{i}$, $\forall i \in \mathcal{O}$, where $\mathcal{O}=\left\{i|i \in \mathbb{Z}_{\left[1,\,n\right]},\, i\neq l \right\}$ with $n$ representing the number of agents, and the corresponding state of the traffic, $s_{t+j}$,  at prediction steps $j=0,\,1,\,\cdots,\,N-1$, where $s_t = \left[x_t{\left[1\right]},\,y_t{\left[1\right]},\,v_t{\left[1\right]},\,\theta_t{\left[1\right]},\,\cdots,\,x_t{\left[n\right]},\,y_t{\left[n\right]},\,v_t{\left[n\right]},\,\theta_t{\left[n\right]}\right]^T$. In this paper, level-k game theory \cite{Costa-Gomes2006,Costa-Gomes2009} is utilized to model the vehicle-to-vehicle interactions and thus predict the actions of the other agents over the horizon.

	In level-$k$ game theory, it is assumed that the decisions taken by the strategic agents are based on the predictions of the actions of the other agents and the agents can have have different reasoning levels. The reasoning depth of an agent is indicated by $k \in \left\{0,\,1,\,\cdots\right\}$. The hierarchy begins with agents at level-0, where the agents make instinctive decisions to achieve the objective without accounting for the interactions between other agents. On the other hand, the agents at level-$k$ $\forall k>0$, consider the interactions by assuming that the other agents are at level-$(k-1)$ and take decisions accordingly. For instance, a \nth{l} level-1 agent assumes other agents are at level-0 and predicts their action sequences, $\actseqk{i}{0}$ $\forall i \in \mathcal{O}$, to compute its own action sequence, $\actseqk{l}{1}$.
	
	The level-k game theory was adapted to model the vehicle-to-vehicle interactions at an unsignalized four-way intersection in \cite{Li2018}. It is assumed that level-0 vehicles consider the other vehicles in the traffic scenario as stationary obstacles. Therefore, these level-0 drivers implicitly assume the others will yield the right of way, and can be regarded `aggressive'. And level-1 drivers consider other drivers to be aggressive and take `cautious' actions. In \cite{Li2018}, the drivers are categorized into level-0, 1 and 2. Since, the behavior of level-2 driver will be similar to that of the level-0 drivers, in this paper, only level-0 and 1 drivers are considered. The stage reward value obtained for \nth{l} level-$k$ agent at a prediction step $j$ for an action $\actk{l}{k}{t+j}$, depend on the current traffic state, $s_0$, the ego agent's actions, $\actseqkt{l}{k}{j-1}$, and the actions of other agents, $\actseqkt{i}{k-1}{j-1}$ $\forall i \in \mathcal{O}$, is given as
	\begin{align}
		R_{t+j}^{(k)}[l] = R_{t+j} & \left( \actk{l}{k}{t+j}  \left| s_0,\, \actseqktnopar{l}{k}{j-1},\right.\right. \nonumber\\
		& \hspace{0.25cm} \left.\,\actseqktnopar{i}{k-1}{j-1}\right),
		\label{eq:stage_reward_multi}
	\end{align}
	
	\noindent and its cumulative reward is
	\begin{align}
		\mathcal{R}^{(k)}\left(\boldsymbol{\gamma_{t}^{(k)}[l]}\right) = \sum_{j=0}^{N-1} \lambda^{j} R_{t+j}^{(k)}[l].
		\label{eq:cum_reward_multi}
	\end{align}

	\subsection{Multi-model strategy}

	Human drivers, initially, do not have perfect knowledge about other drivers. However, they gain better understandings of other driver's characteristics through interactions, and therefore, resolve conflicts effectively. Similarly, the autonomous vehicles in a multi-agent traffic scenario, hold an initial belief about the driver model (levels-0 and 1) of the other vehicles as a probability distribution over both models. This facilitates the expression of a driver's degree of aggressiveness (or cautiousness) as a continuous parameter between levels-0 and 1.  Subsequently, based on the actual action applied by the other agents, the estimate of the probability distribution is updated at every step.
	
	From the perspective of an \nth{l} autonomous agent, the probability that the \nth{i} other agent can be modeled as level-$k$ is represented by $P_{K_{i}^l = k}$. The probability of the model $k$ is increased when it matches the actual action by 
	\begin{subequations} \label{lv_est}
		\begin{align}
			&k^* = \arg \, \underset{k \in \left\{0,\,1 \right\}} {\min} \left\| \actt{i}{k}{t} - \actk{i}{k}{t}  \right\|  \\ 
			& \tilde{P}_{K_{i}^l = k^*} \left(t\right) = {P}_{K_{i}^l = k^*} \left(t-1\right) + \Delta P  \\ 
			& {P}_{K_{i}^l = k}\left(t\right) = \frac{\tilde{P}_{K_{i}^l = k}\left(t\right) }{\sum_{\tilde{k} = 0}^{1} \tilde{P}_{K_{i}^l = \tilde{k}}\left(t\right) },  \forall k \in \left\{0,\,1 \right\},
		\end{align}
		\label{eq:model_update}
	\end{subequations}
	
	\noindent where $\actt{i}{k}{t}$ and $\actk{i}{k}{t}$ represent the actual and predicted action taken by \nth{i} agent assuming level-$k$ model, respectively; $\Delta P >0 $ is a constant that denotes the rate of increment of the probability;  $\left\|\actt{i}{k}{t} - \actk{i}{k}{t}\right\| = \left| a\left(t\right) - a_t^{(k)} \right| + \left|  \delta_f\left(t\right) - \delta_{f,\,t}^{(k)} \right|$. When the input pair of the actual action is equal to that of the predicted action, the probability distribution remains unchanged.

	In order to incorporate the multi-model strategy in the decision making process and select the optimal action according to the model of other agents, the expected cumulative reward for the \nth{l} agent, using \eqref{eq:stage_reward_multi} and \eqref{eq:cum_reward_multi}, is given by
	\begin{align}
		\mathcal{R}_P\left(\boldsymbol{\gamma_{t}}[l]\right)  = \sum_{k=0}^{1} {P}_{K_{i}^l = k}\left(t\right)  \mathcal{R}^{(k)}\left(\boldsymbol{\gamma_{t}^{(k)}}[l]\right),\, \forall i \in \mathcal{O}.
		\label{eq:cum_reward_prob}
	\end{align}

	\subsection{Robust decision making}
	\label{sec:robust}

	The mismatch between the actual position of the center of mass of the vehicle which is used to determine the rectangular outer approximation of the vehicle (see Section \ref{sec:reward_function}) and its predictions obtained using the dynamic model in \eqref{eq:nom_mod} might lead to collision. In the multi-agent traffic scenario under consideration, there are two sources of modeling errors: (i) the uncertainties, $w^{(m)} = \left(w_x^{(m)},\,w_y^{(m)}\right) \in \mathcal{W}_m$, resulting due to the use of a simplified model in \eqref{eq:nom_mod}; and (ii) the uncertainty, $w^{(d)} = \left(w_x^{(d)},\,w_y^{(d)}\right) \in \mathcal{W}_d$, arising due to unknown driver model. Hence, the disturbance set defined in \eqref{eq:dist_set} is 
	\begin{align}
		\mathcal{W} = \mathcal{W}_m \oplus \mathcal{W}_d.
		\label{eq:full_dist_set}	
	\end{align}

	Robust approaches can be used to account for these uncertainties while computing the control actions. Since a discrete set of input actions is considered in this work, feedback min-max strategy \cite{Scokaert1998} is utilized in this work for considering the uncertainties originating from the disturbance set $\mathcal{W}$. Since the autonomous agents update the driver models of the other agents in the multi-agent traffic scenario at each step according to \eqref{eq:model_update}, an adaptive scheme is proposed to incorporate the confidence on the driver models and leverage the fact that level-1 drivers are cautious. The disturbance set considered at each time step is modified  of the other agents and leverage .

	\begin{align}
		\bar{\mathcal{W}}_{i}^{l}(t) = \mathcal{W}_m \oplus P_{K_{i}^l = 0}(t) \mathcal{W}_d
		\label{eq:adaptive_dist_set}
	\end{align}

	\noindent where at time $t$, $P_{K_{i}^l = 0} (t) $ is the probability that the \nth{i} agent is a level-0 driver from the perspective of the \nth{l} autonomous agent, and $\bar{\mathcal{W}}_{i}^{l}(t)$ denotes the disturbance set. It is assumed that, initially, all the agents are level-0 drivers, i.e., $P_{K_{i}^l = 0} (0) = 1,\, \forall i \in \mathcal{O}$. Essentially, this assumption allows the autonomous agent to be cautious with another interacting agent when there is no/less information about that agent. If an \nth{i} agent is level-0, as time evolves, $P_{K_{i}^l = 0} (t)$ will continue to be equal to one, and hence, the autonomous agent take conservative actions (or behave like level-1 driver). On the other hand, when the \nth{i} agent is a level-1 driver, $P_{K_{i}^l = 0} (t)$ will decrease, resulting in a reduced disturbance set size, thereby, allowing the autonomous agent to take less conservative actions and adapt to the behavior of the interacting agents while capable of handling the uncertainties arising due to the use of a simple prediction model. 

	When interacting with an agent $i\in \mathcal{O}$, the objective the autonomous agent $l$ is to maximize the expected cumulative reward \eqref{eq:cum_reward_multi}, while accounting for the effect of the worst-case uncertainty from the possible disturbance realizations. The optimal control sequence, $\optactseq{l}$, is obtained by solving the following optimization problem 
	\begin{subequations}
		\begin{align}
			\boldsymbol{\gamma^{*}_{t}[l]} =  & \arg \underset {\boldsymbol{\gamma_{t}[l]} \in  \boldsymbol{\Gamma}} {\max}\, \, \underset{w_{t+j}^{p} \in \bar{\mathcal{W}}_{i}^{l}(t)}{\min} \, \, \mathcal{R}_P\left(\boldsymbol{\gamma_{t}[l]}\right) \\
			\textrm{s. t.} \forall &  j \in \mathbb{Z}_{0,\,N-1},\, \eqref{eq:psi_pred},\, \eqref{eq:v_pred},\, \eqref{eq:beta_pred} ,\,\forall i \in \mathcal{O},\, \forall p \in \mathcal{P} \nonumber \\
			\tilde{x}_{t+j+1} & = 	\tilde{x}_{t+j} + v_{t+j} \cos\left(\psi_{t+j} + \beta_{t+j} \right) \Delta t + w_{x,\,t+j}^{p}\\ 
			\tilde{y}_{t+j+1} & = 	\tilde{y}_{t+j} + v_{t+j} \sin\left(\psi_{t+j} + \beta_{t+j} \right) \Delta t + w_{x,\,t+j}^{p},
			\label{eq:control_action_model}
		\end{align}
		\label{eq:control_prob}
	\end{subequations}

	\noindent where $w_{t+j}^{p} = \left(w_{x,\,t+j}^{p},\, w_{y,\,t+j}^{p}\right)$ denote a possible realization of the uncertainty in the global position of the center of mass in $x$ and $y$ directions, respectively; and $\mathcal{P}$ represents the set of indexes the realizations. Since the decision variables of \eqref{eq:control_prob} are discrete, it is solved using a decision tree approach. The autonomous agent then applies the first element $\optact{l}$ of the optimal control action sequence, i.e.,  $\gamma\left(t\right) = \optact{l}$.

	
	
	\begin{figure*}
		\begin{centering}
			\begin{tikzpicture}[scale=0.4,transform shape]
			\node (origin) at (0,0) {};
			
			\node (adp1)[below left = 0cm and 0cm of origin, minimum width = 0.cm, minimum height = 0cm, inner sep = 0pt]{\includegraphics[clip, trim = {1.5cm 0.25cm 1.5cm 0.25cm}]{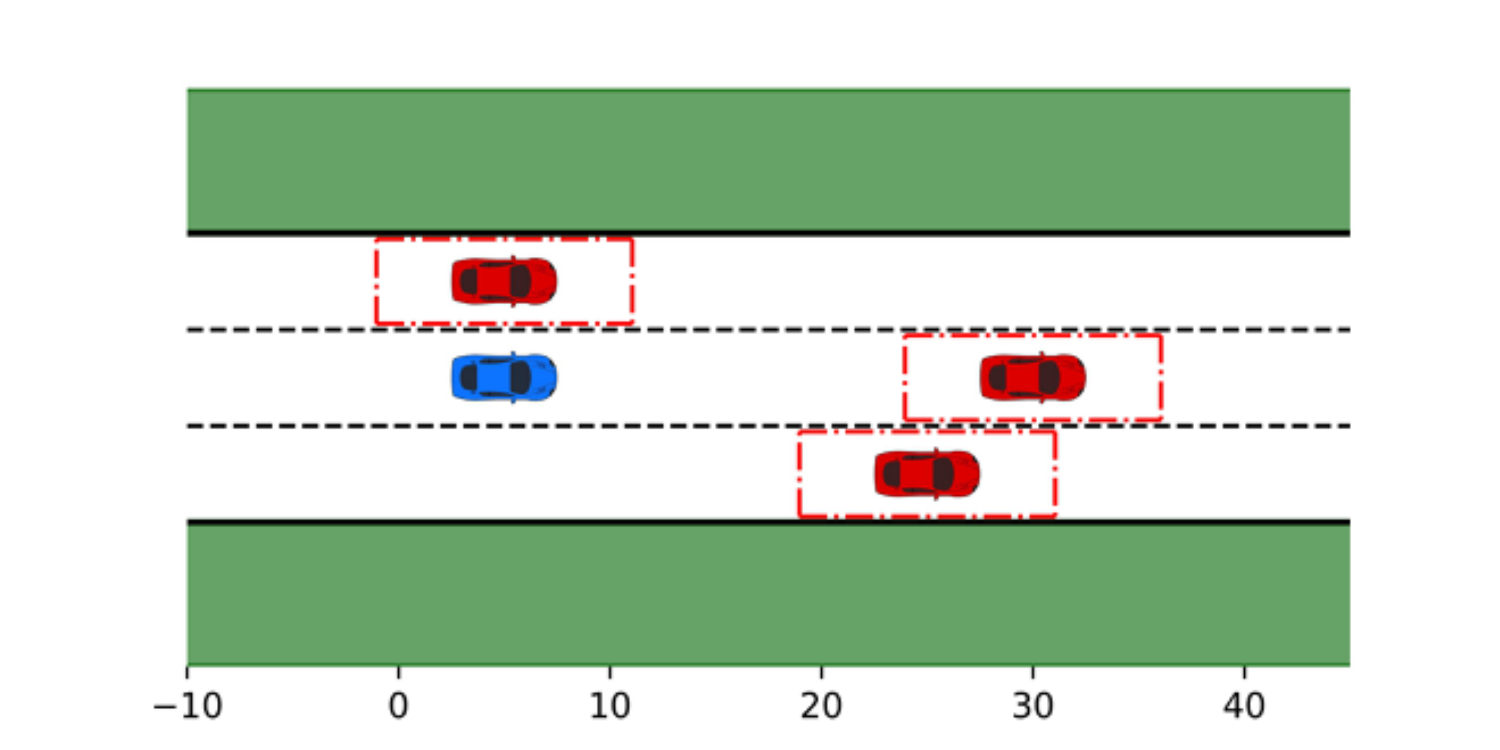}};
			\node (agg1)[left = 1cm of adp1, minimum width = 0.cm, minimum height = 0cm, inner sep = 0pt]{\includegraphics[clip, trim = {1.5cm 0.25cm 1.5cm 0.25cm}]{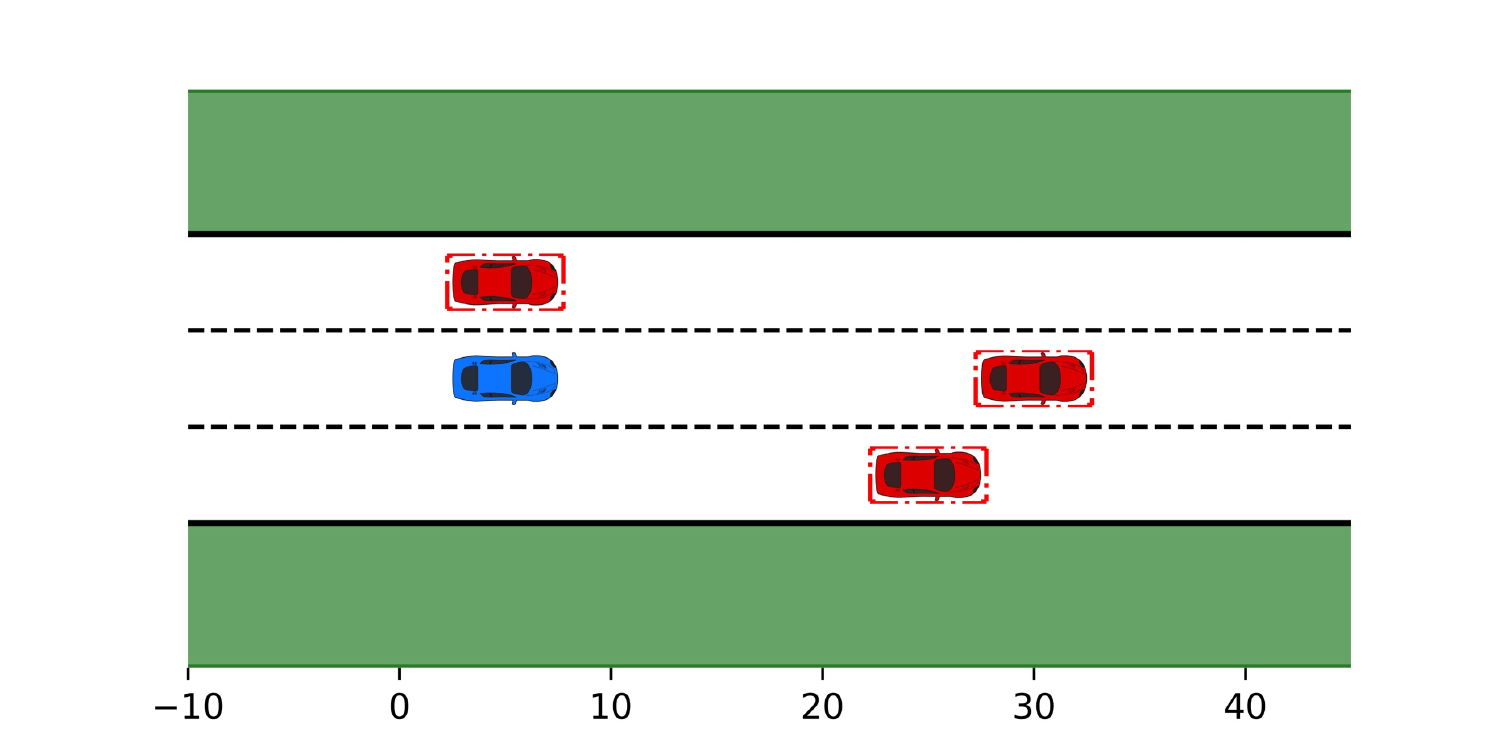}};
			\node (con1)[ right = 1cm of adp1, minimum width = 0.cm, minimum height = 0cm, inner sep = 0pt]{\includegraphics[clip, trim = {1.5cm 0.25cm 1.5cm 0.25cm}]{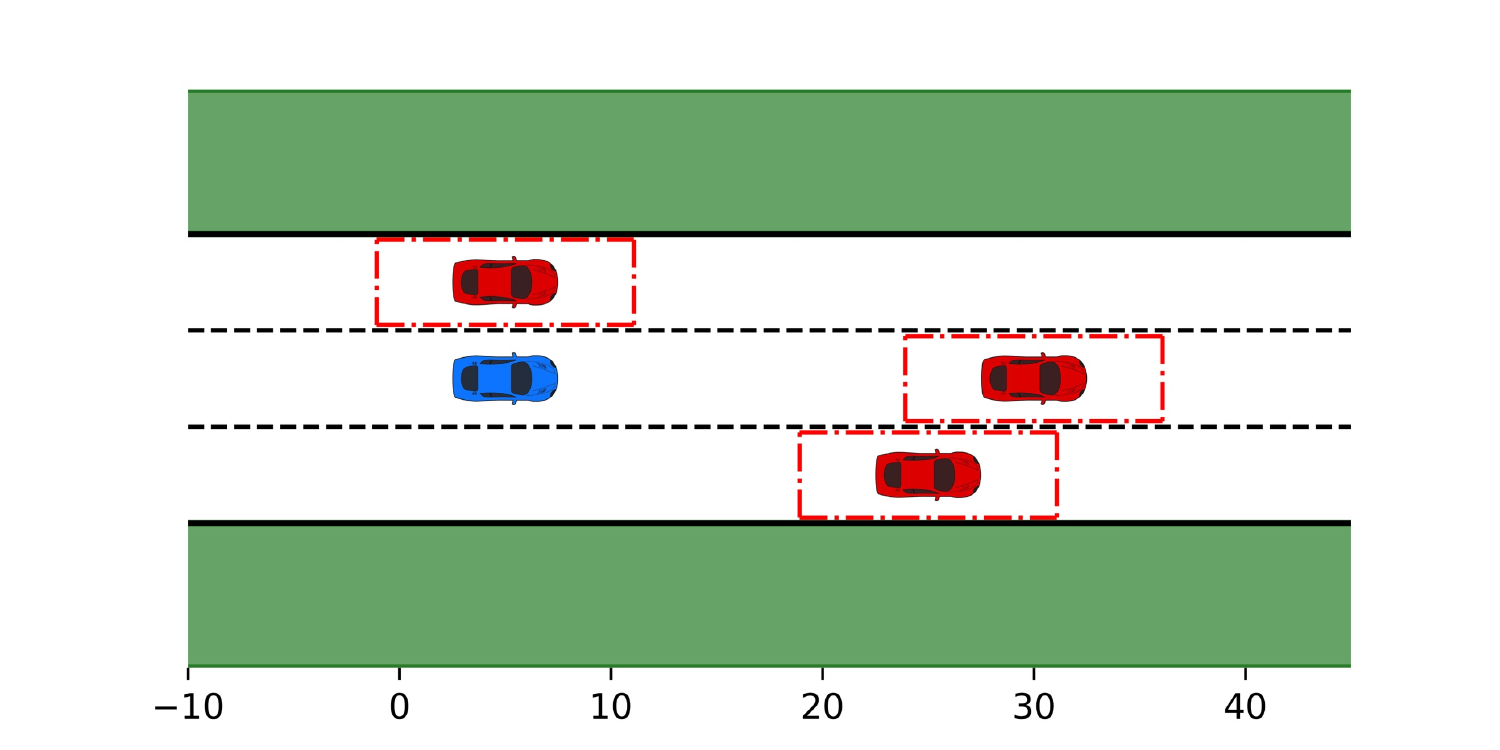}};

			\node (adp2)[below  = 1.cm of adp1, minimum width = 0.cm, minimum height = 0cm, inner sep = 0pt]{\includegraphics[clip, trim = {1.5cm 0.25cm 1.5cm 0.25cm}]{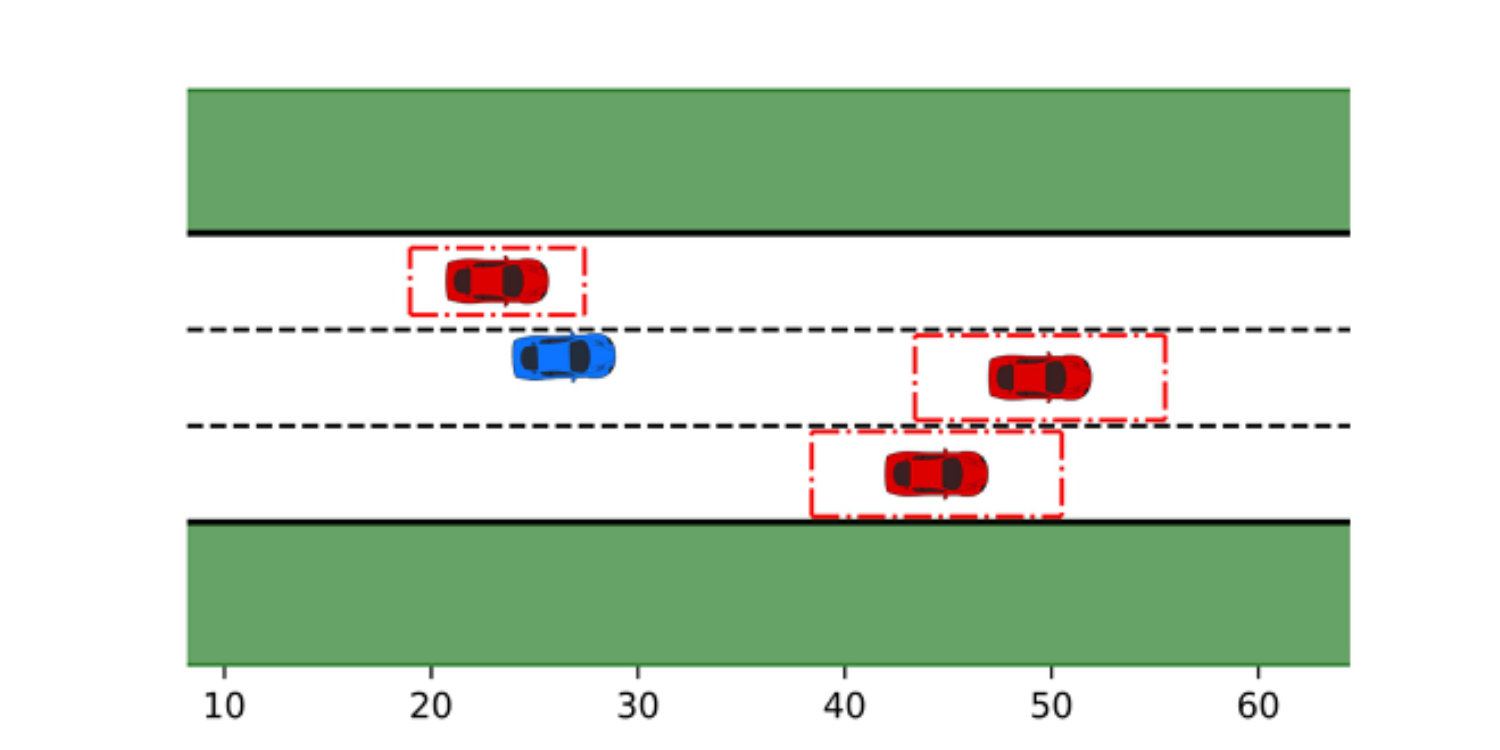}};
			\node (agg2)[left = 1cm of adp2, minimum width = 0.cm, minimum height = 0cm, inner sep = 0pt]{\includegraphics[clip, trim = {1.5cm 0.25cm 1.5cm 0.25cm}]{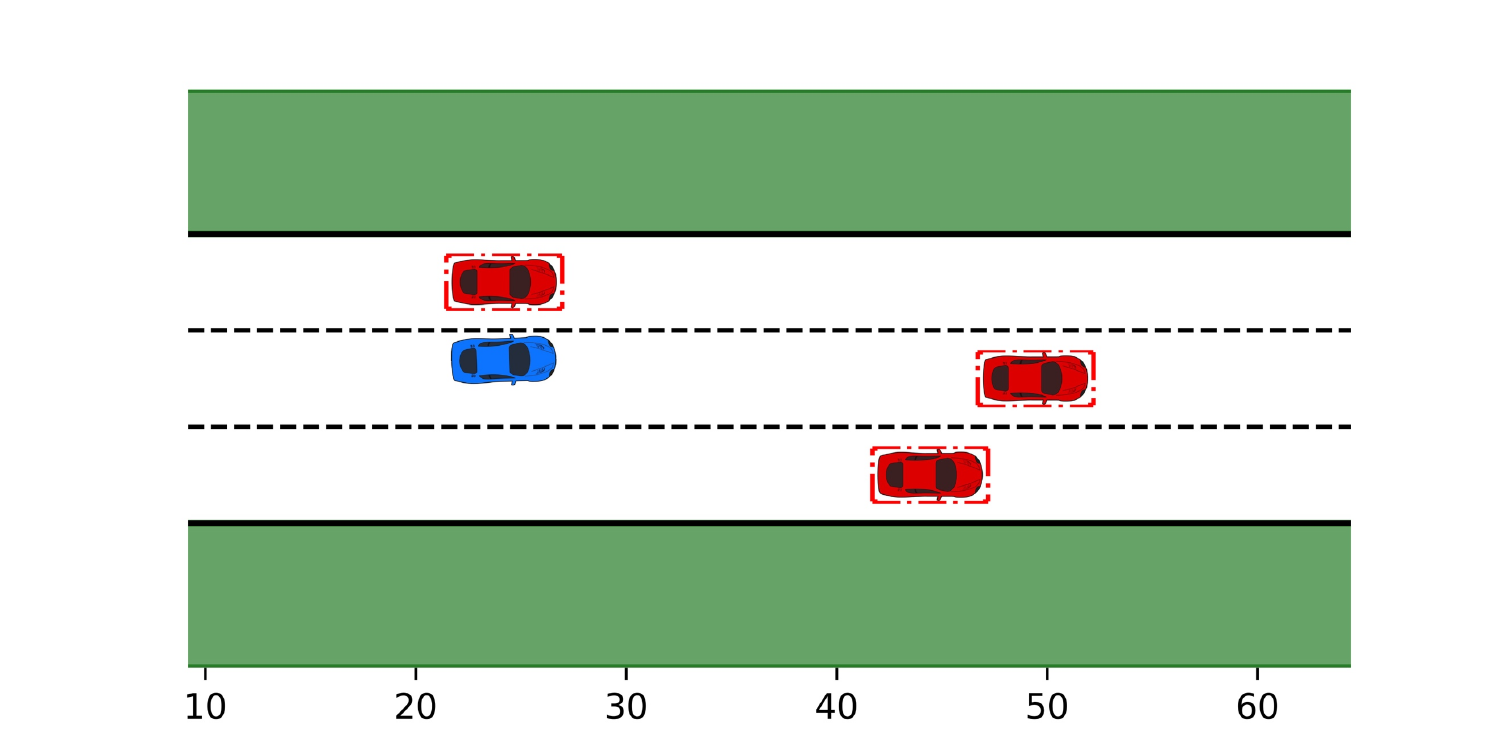}};
			\node (con2)[ right = 1cm of adp2, minimum width = 0.cm, minimum height = 0cm, inner sep = 0pt]{\includegraphics[clip, trim = {1.5cm 0.25cm 1.5cm 0.25cm}]{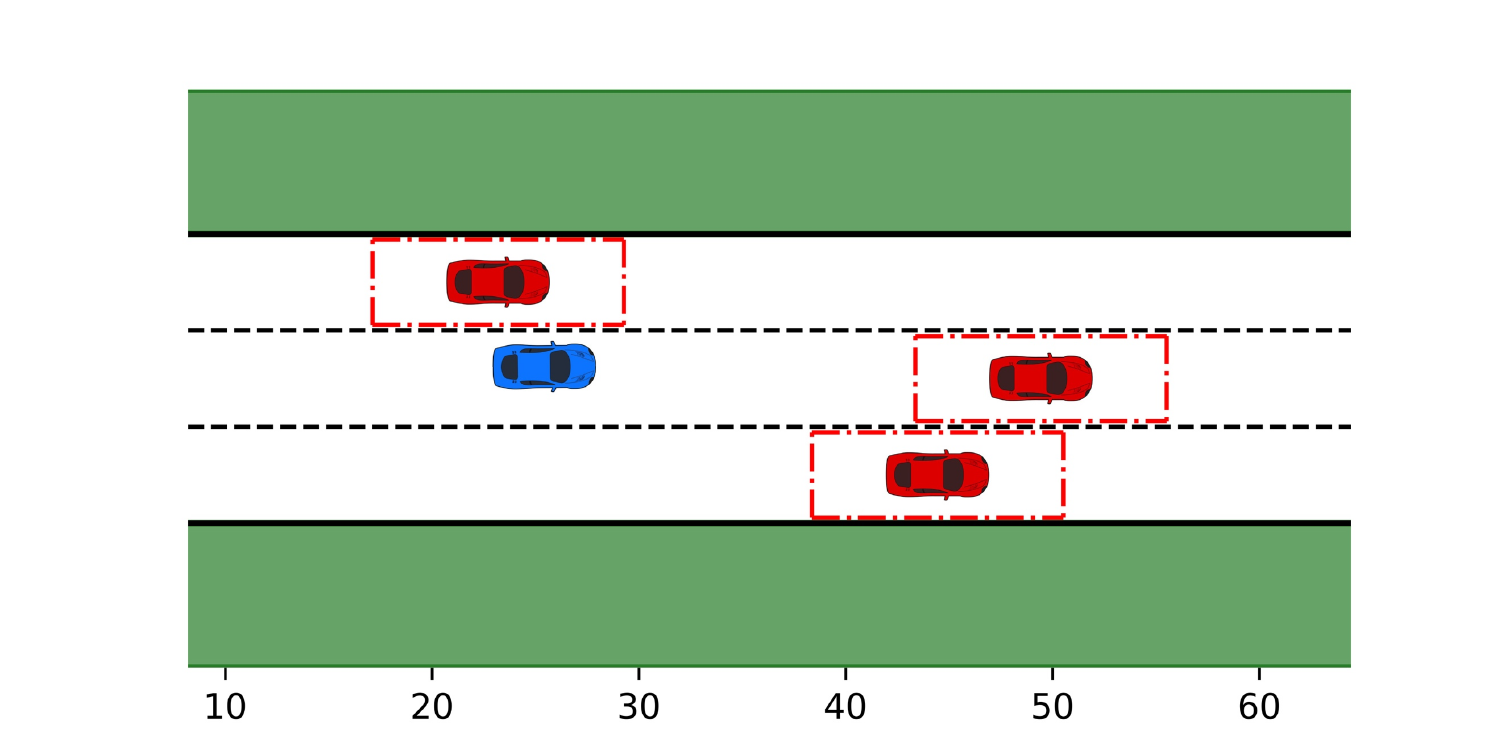}};

			\node (adp3)[below  = 1.cm of adp2, minimum width = 0.cm, minimum height = 0cm, inner sep = 0pt]{\includegraphics[clip, trim = {1.5cm 0.25cm 1.5cm 0.25cm}]{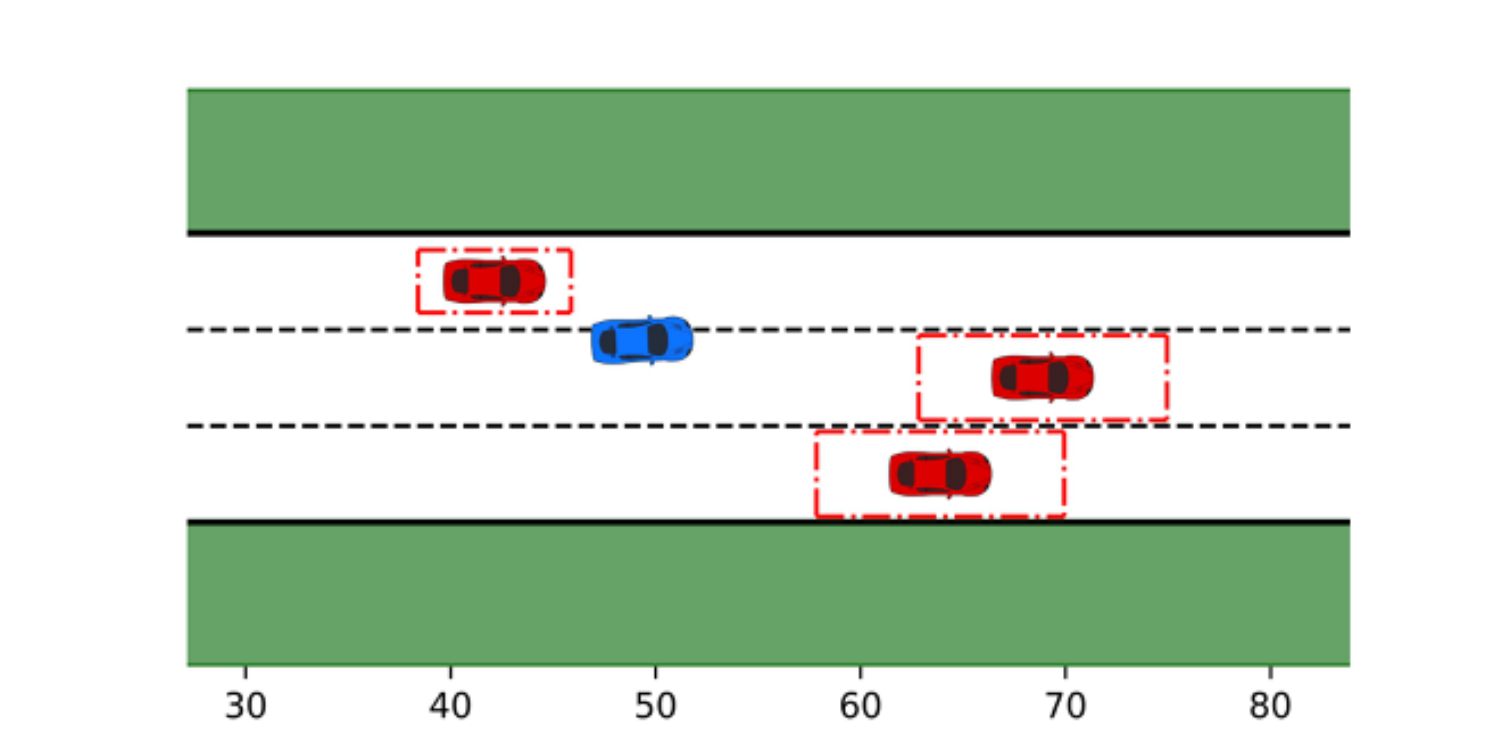}};
			\node (agg3)[left = 1cm of adp3, minimum width = 0.cm, minimum height = 0cm, inner sep = 0pt]{\includegraphics[clip, trim = {1.5cm 0.25cm 1.5cm 0.25cm}]{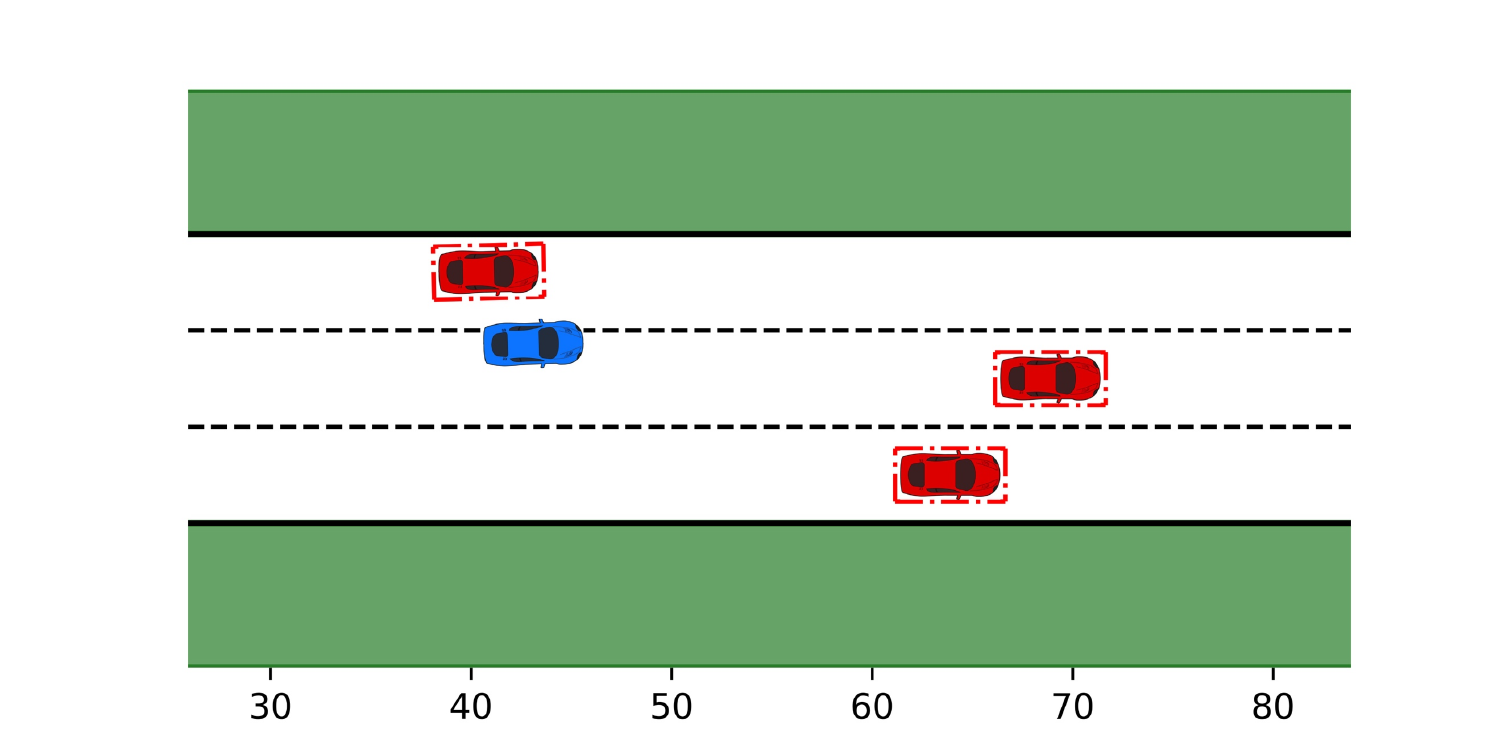}};
			\node (con3)[ right = 1cm of adp3, minimum width = 0.cm, minimum height = 0cm, inner sep = 0pt]{\includegraphics[clip, trim = {1.5cm 0.25cm 1.5cm 0.25cm}]{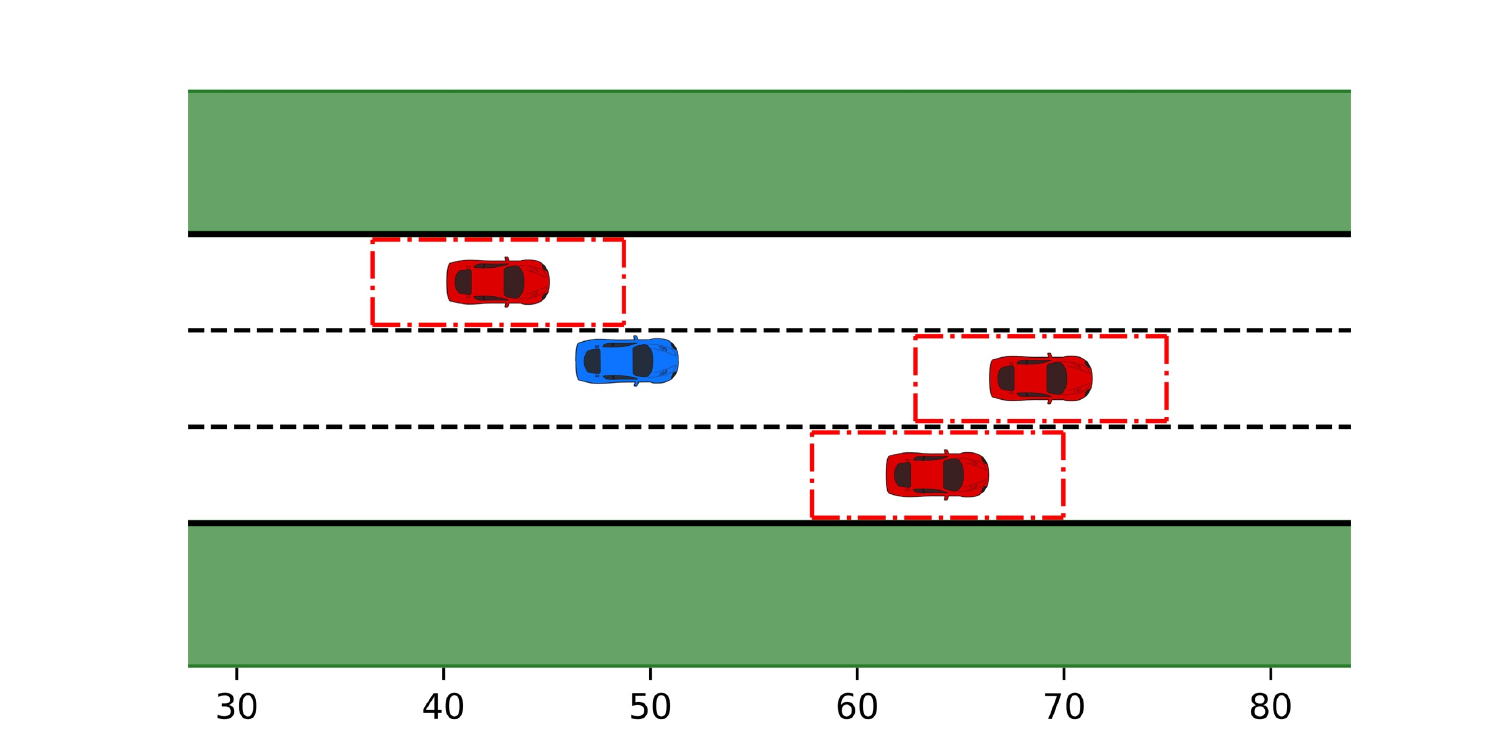}};

			\node (adp4)[below  = 1.cm of adp3, minimum width = 0.cm, minimum height = 0cm, inner sep = 0pt]{\includegraphics[clip, trim = {1.5cm 0.25cm 1.5cm 0.25cm}]{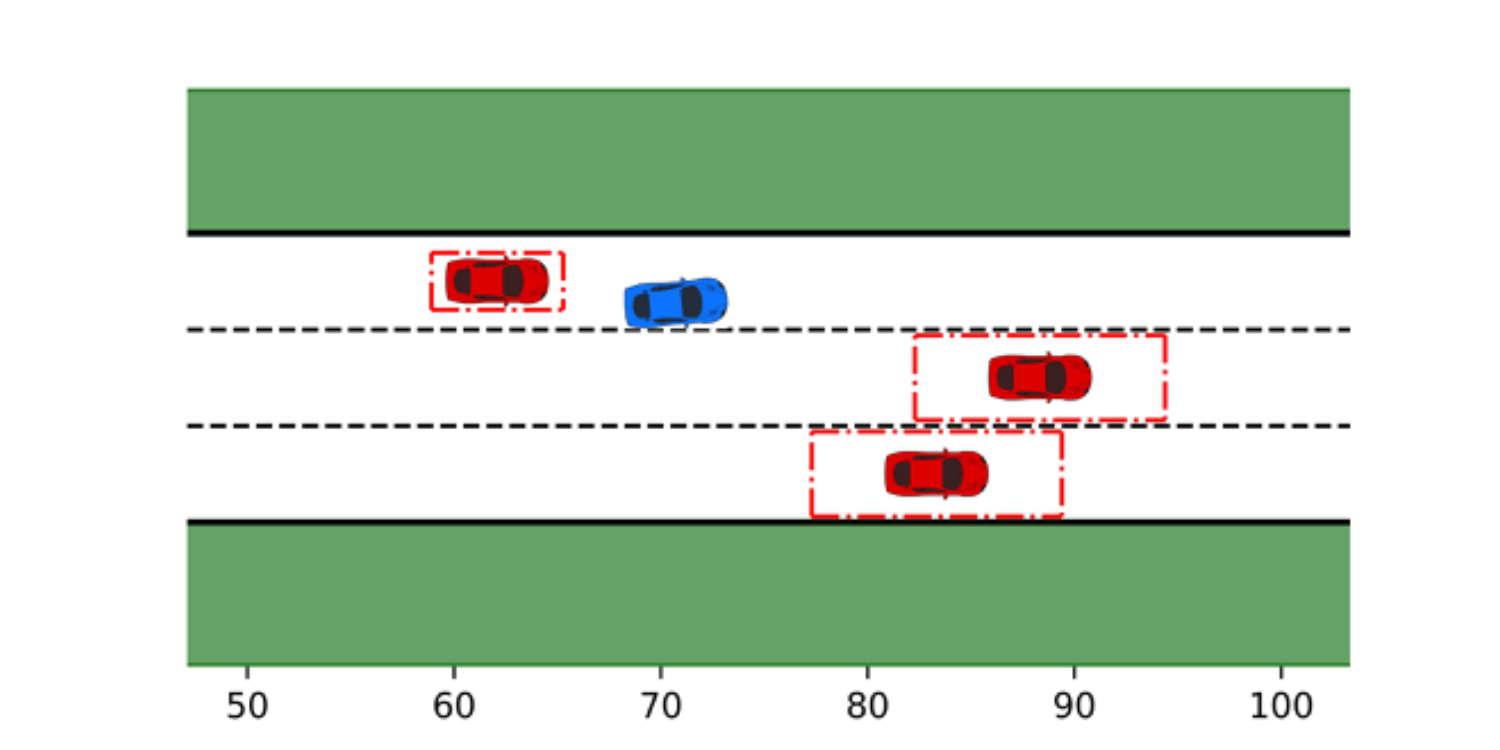}};
			\node (agg4)[left = 1cm of adp4, minimum width = 0.cm, minimum height = 0cm, inner sep = 0pt]{\includegraphics[clip, trim = {1.5cm 0.25cm 1.5cm 0.25cm}]{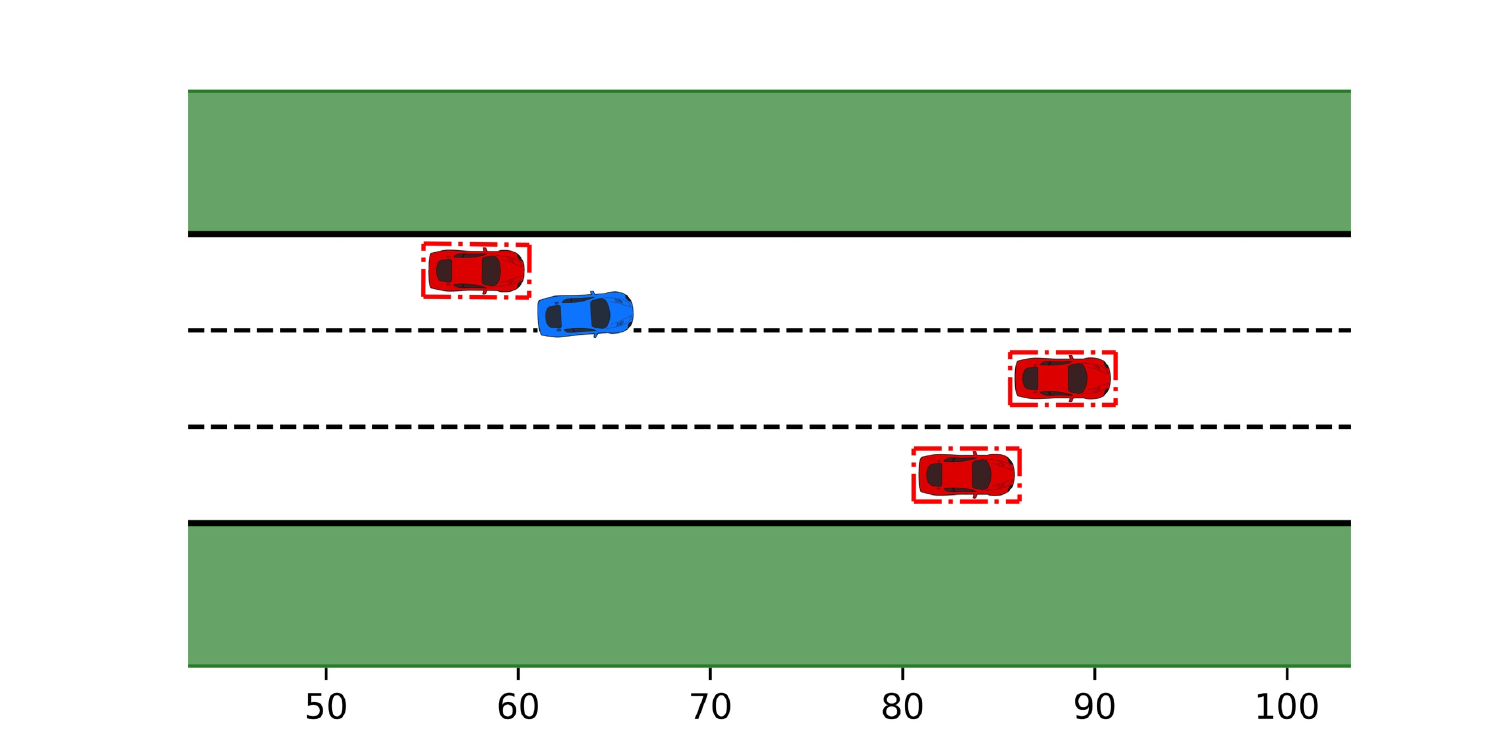}};
			\node (con4)[ right = 1cm of adp4, minimum width = 0.cm, minimum height = 0cm, inner sep = 0pt]{\includegraphics[clip, trim = {1.5cm 0.25cm 1.5cm 0.25cm}]{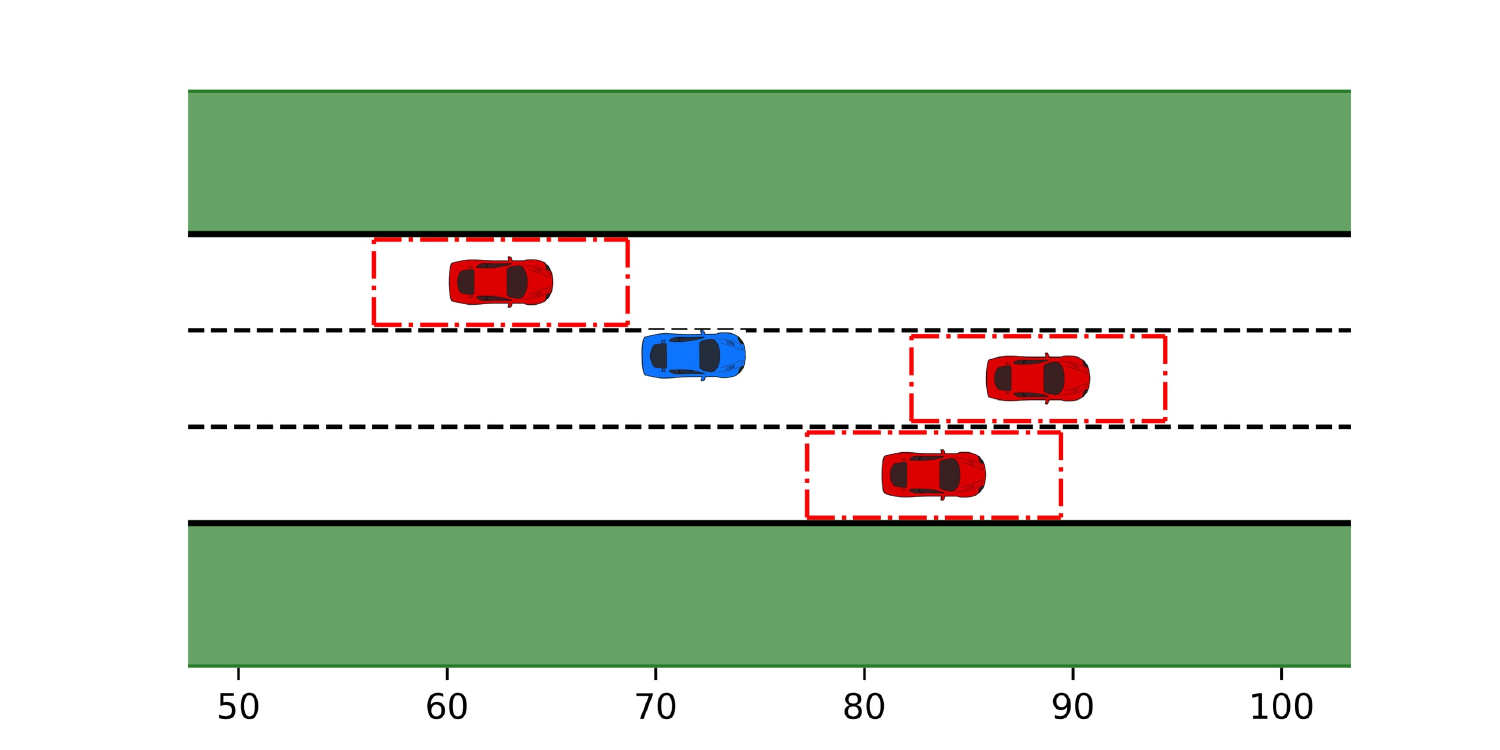}};
			
			\node (adp5)[below  = 1.cm of adp4, minimum width = 0.cm, minimum height = 0cm, inner sep = 0pt]{\includegraphics[clip, trim = {1.5cm 0.25cm 1.5cm 0.25cm}]{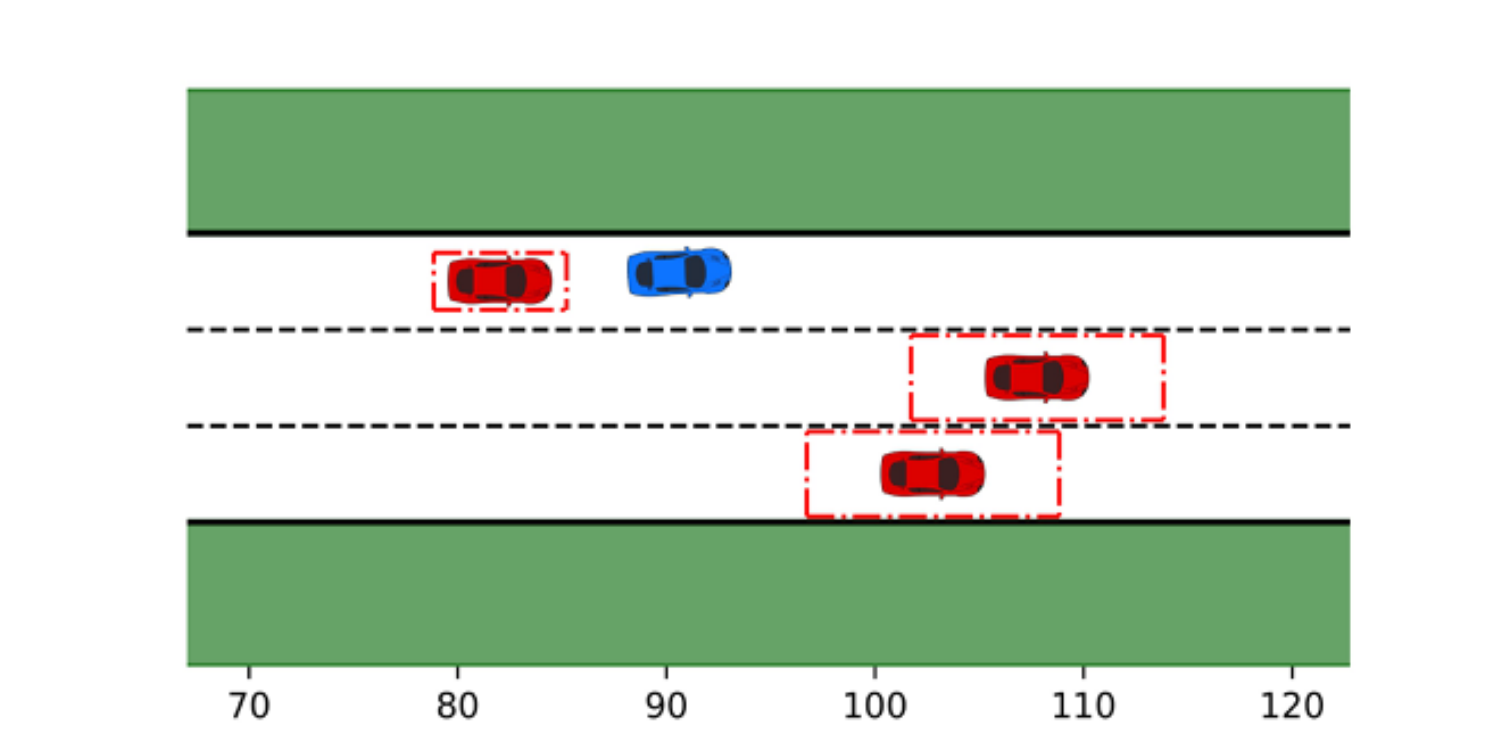}};
			\node (agg5)[left = 1cm of adp5, minimum width = 0.cm, minimum height = 0cm, inner sep = 0pt]{\includegraphics[clip, trim = {1.5cm 0.25cm 1.5cm 0.25cm}]{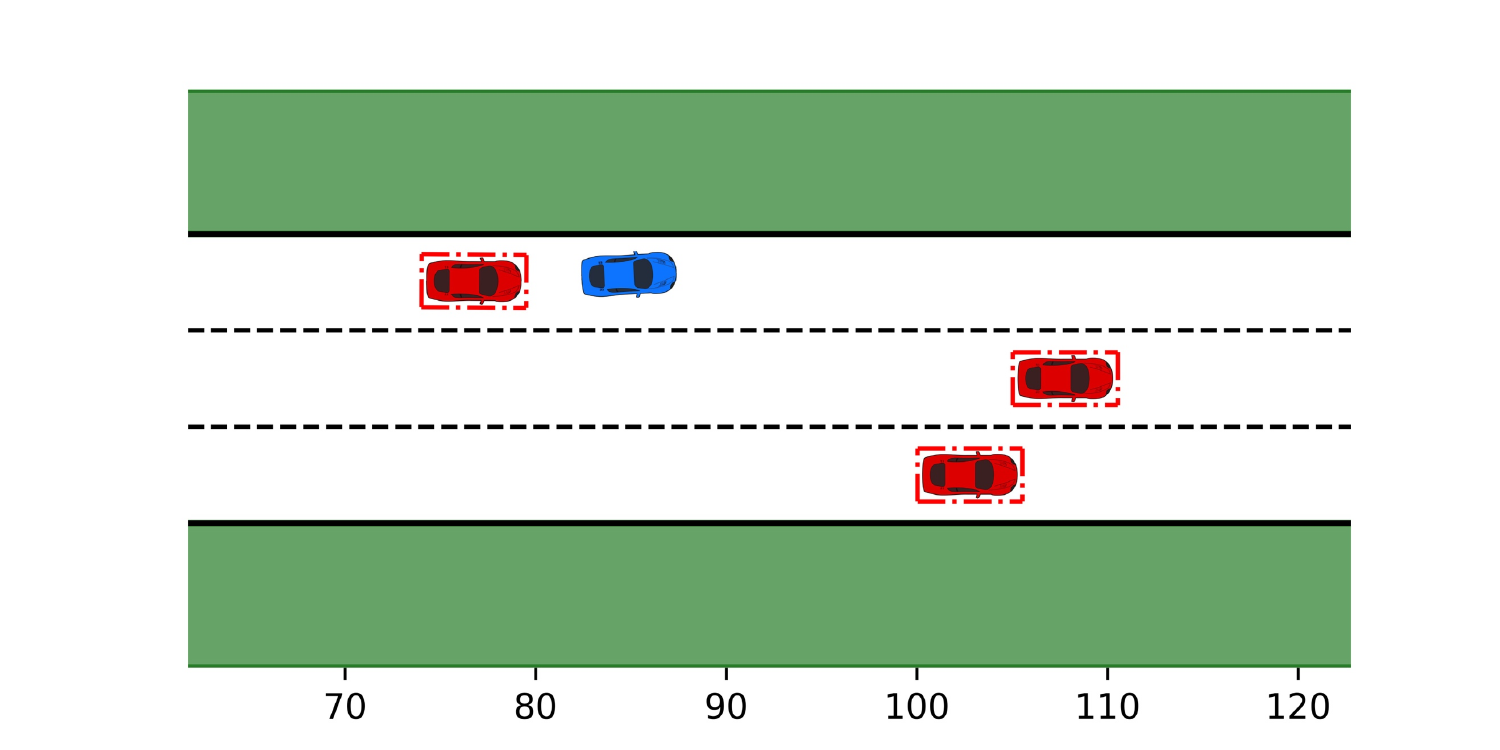}};
			\node (con5)[ right = 1cm of adp5, minimum width = 0.cm, minimum height = 0cm, inner sep = 0pt]{\includegraphics[clip, trim = {1.5cm 0.25cm 1.5cm 0.25cm}]{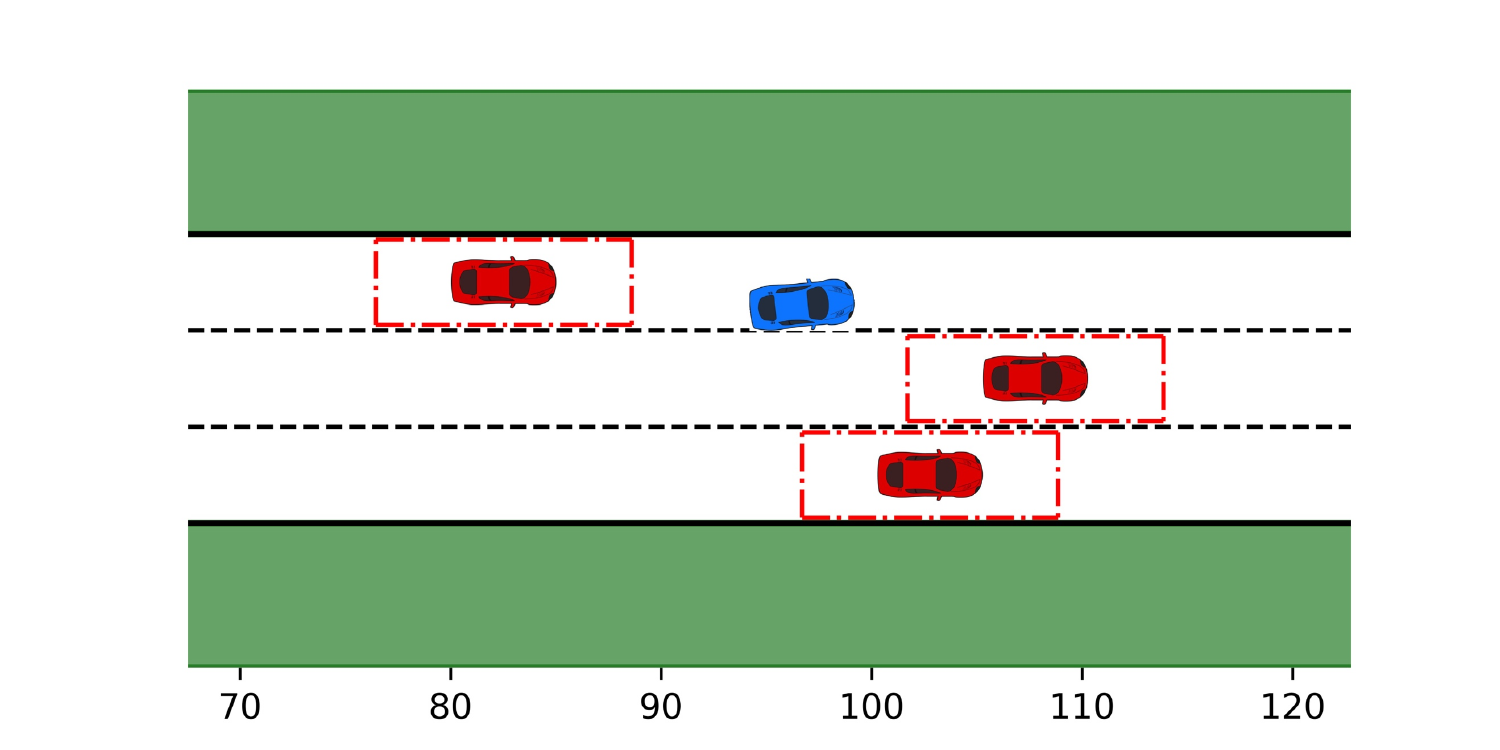}};
			%

			\node (xlabeladp) [below = 1cm of adp5, inner sep = 0pt] {\Huge Distance (m)};
			\node (xlabelagg) [below = 1cm of agg5, inner sep = 0pt] {\Huge Distance (m)};
			\node (xlabelcon) [below = 1cm of con5, inner sep = 0pt] {\Huge Distance (m)};

			\node(car1) [below left = -2.79cm and -6.75cm of agg1, inner sep = 0pt, draw, circle, minimum height = 0.7cm, minimum width = 0.7cm] {\LARGE 1};
			\node(car2) [below left = -3.79cm and -2.45cm of agg1, inner sep = 0pt, draw, circle, minimum height = 0.7cm, minimum width = 0.7cm] {\LARGE 2};
			\node(car3) [below left = -3.79cm and -7.95cm of agg1, inner sep = 0pt, draw, circle, minimum height = 0.7cm, minimum width = 0.7cm] {\LARGE 3};
			\node(car4) [below left = -4.79cm and -2.45cm of agg1, inner sep = 0pt, draw, circle, minimum height = 0.7cm, minimum width = 0.7cm] {\LARGE 4};

			\node(lane1) [below left = -2.76cm and -11.85cm of agg1, inner sep = 0pt] {\LARGE I};
			\node(lane1) [below left = -3.76cm and -11.95cm of agg1, inner sep = 0pt] {\LARGE II};
			\node(lane1) [below left = -4.76cm and -12.05cm of agg1, inner sep = 0pt] {\LARGE III};
			
			\node (t0) [left = 0.25cm of agg1, inner sep = 0pt] {\Huge $t=\SI{0}{s}$};
			\node (t1) [left = 0.25cm of agg2, inner sep = 0pt] {\Huge $t=\SI{1}{s}$};
			\node (t2) [left = 0.25cm of agg3, inner sep = 0pt] {\Huge $t=\SI{2}{s}$};
			\node (t3) [left = 0.25cm of agg4, inner sep = 0pt] {\Huge $t=\SI{3}{s}$};
			\node (t4) [left = 0.25cm of agg5, inner sep = 0pt] {\Huge $t=\SI{4}{s}$};
			
			\node (agg) [above = 0.25cm of agg1, inner sep = 0pt] {\Huge (a) Nominal};
			\node (adp) [above = 0.25cm of adp1, inner sep = 0pt] {\Huge (b) Adaptive};
			\node (con) [above = 0.25cm of con1, inner sep = 0pt] {\Huge (c) Robust};

			\end{tikzpicture}
			\par\end{centering}
		\protect\caption{A four second simulation sequence with a one second time interval (see along the column) showing the lane changing maneuver performed by the autonomous vehicle under  (a) nominal; (b) adaptive; and (c) robust decision making strategies in a multi-traffic scenario. The circled numbers indicate the vehicle ids and the Roman numerals denote the lane number in the subfigure $[1,\,1]$. The dash-dotted lines indicate the set $s \oplus \bar{\mathcal{W}}_{i}^{l}(t),\, \forall i \in \mathcal{O}$ from the perspective of the autonomous agent.}
		\label{fig:snapshots}
	\end{figure*}
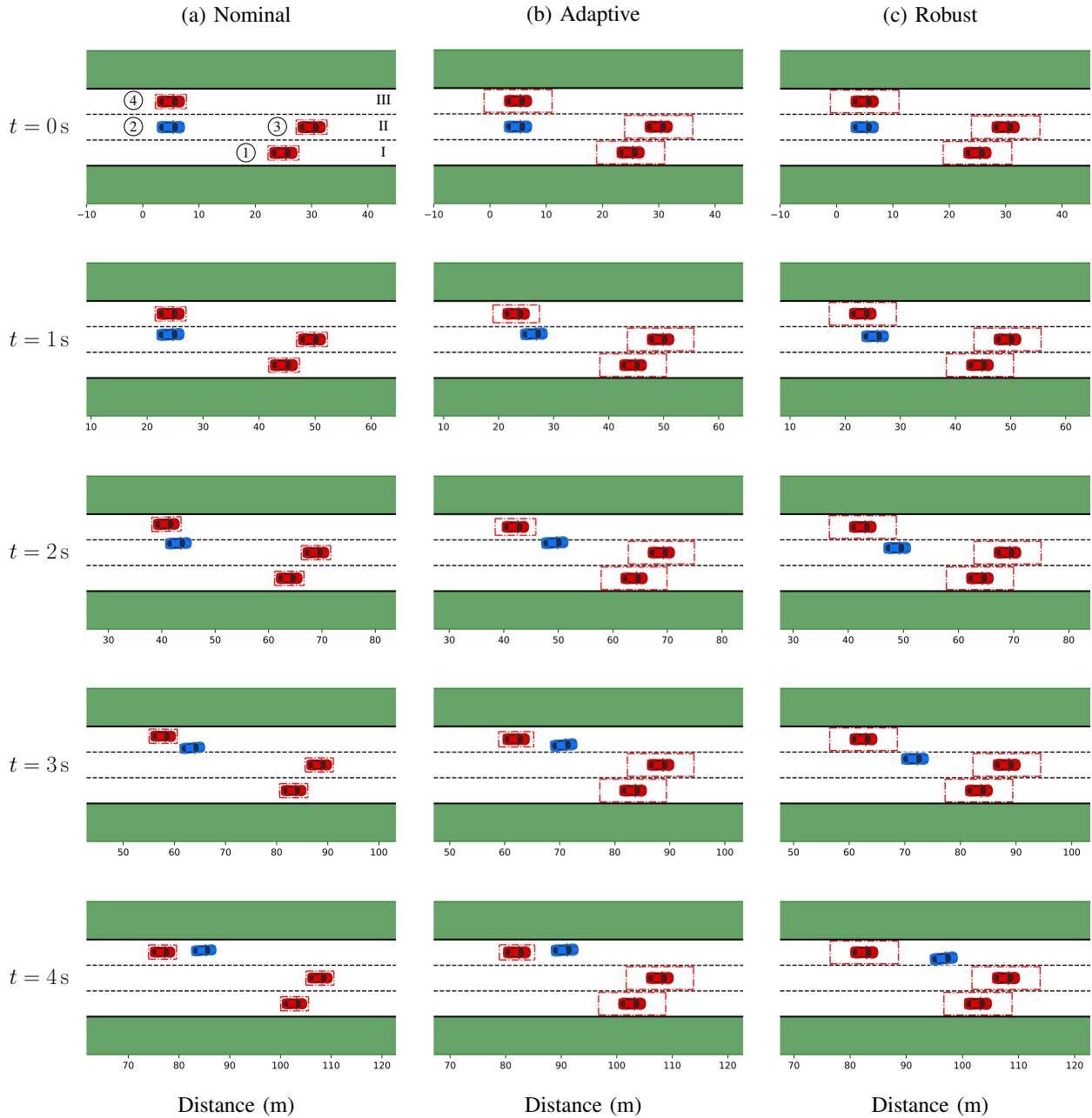

	
	\section{Simulation results}
	\label{sec:sim_results}
	The proposed adaptive robust approach is validated for the lane changing maneuver on a numerically simulated three lane highway section\footnote{The code is made publicly available at \url{https://github.com/gokulsivasankar/RobustDecisionMaking}}. Consider the multi-agent traffic scenario shown in the subfigure $[1,\, 1]$ (first element represents row and second element represents column) in Fig.~\ref{fig:snapshots}. The objective of the autonomous agent (blue) is to change from lane II to lane III, while the human agents keep their respective lane. All the human drivers are assumed to be level-1, i.e., they exhibit cautious behavior, however, it is unknown to the AV. The sampling time is set to \SI{0.5}{s} and two step prediction horizon is considered. The proposed methodology is compared to the nominal, and the robust strategies during the decision making process. 

	The traffic simulation under the nominal decision making strategy is shown in the subfigures in the first column of Fig.~\ref{fig:snapshots}. In this case, the disturbance set considered for solving \eqref{eq:control_prob} is chosen as an empty set, i.e., $\bar{\mathcal{W}}_{i}^{l}(t) = \emptyset$. Since the disturbances are not considered in this case, it can be noted the AV chooses to steer left as soon as the simulation begins which provides the maximum reward. This move is considered to be aggressive. However, the human vehicle $4$, being cautious, reacts by steering left at time $t = \SI{2}{s}$ and returning to the lane center at  $t = \SI{4}{s}$ once the AV has passed. The AV completes the lane change between \SI{60}{m} and \SI{70}{m}. 

	\begin{figure}
		\begin{centering}
			\begin{tikzpicture}[scale=0.5,transform shape]
			\node (origin) at (0,0) {};
			
            
            \node (fig)[below left = 0cm and 0cm of origin, minimum width = 0.cm, minimum height = 0cm, inner sep = 0pt]{\includegraphics[clip, trim = {1.25cm 0.8cm 1.5cm 1.25cm}]{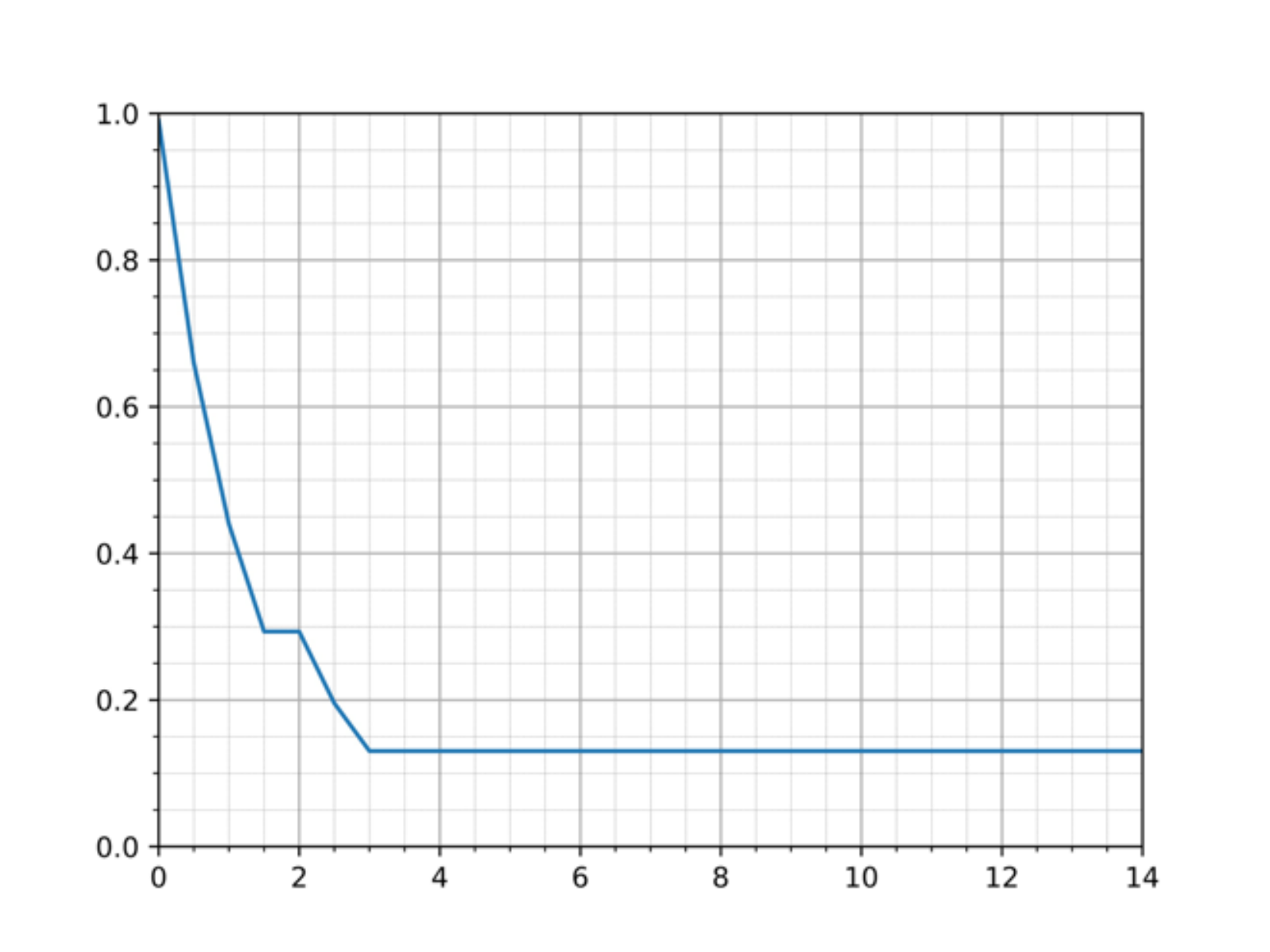}};

			\node (xlabel) [below = 0.5cm of fig, inner sep = 0pt] {\LARGE Time (s)};
		
			\node (ylabel) [below left = -6.1cm and 1cm of fig, inner sep = 0pt, rotate = 90] {\huge $P_{K_{4}^2 = 0}$};
			
			\end{tikzpicture}
			\par\end{centering}
		\protect\caption{Probability that the human driven vehicle 4 can be modeled as level-0 from the perspective of the autonomous vehicle 2 reduces as time evolves since vehicle 4 is level-1.}
		\label{fig:level_history}
	\end{figure}
	
	The results obtained by using the robust strategy that considers $\bar{\mathcal{W}}_{i}^{l}(t) = \mathcal{W}$, for the decision making process, is shown in the subfigures in the third column of Fig.~\ref{fig:snapshots}. The disturbance set $\mathcal{W}$ is defined as in \eqref{eq:full_dist_set}. As mentioned in Section \ref{sec:robust}, initially, all the human drivers are considered to be level-0 from the perspective of the AV, while they are assigned to be level-1 drivers. As seen in Fig.~\ref{fig:level_history}, due to interactions with vehicle $4$, as time evolves, the AV is able to reduce the probability of vehicle $4$ being level-0. It can also be noted from subfigure $[3,\,3]$ in Fig.~\ref{fig:snapshots}, robust control actions are taken to account for the set, $s \oplus \bar{\mathcal{W}}_{i}^{l}(t)$, dash-dotted box surrounding the human drivers shown in the Fig.~\ref{fig:snapshots}. As a result, the AV is conservative in performing the lane change maneuver by completing it between \SI{90}{m} and \SI{100}{m}. 
	
	Lastly, the simulation results for the adaptive control strategy is shown in the subfigures in the second column of Fig.~\ref{fig:snapshots}. The initial disturbance set of all the human agents are similar to the previous case. However, by using the adaptive disturbance set in \eqref{eq:adaptive_dist_set} for computing the control actions, the AV is able to complete the lane change around \SI{70}{m}. Also, it can be observed that the size of the set, $s \oplus \bar{\mathcal{W}}_{i}^{l}(t), \, \forall i \in \{1,\,3\}$, remains constant because, the level-0 and 1 actions are the same for these two vehicles and therefore, according to \eqref{eq:model_update}, the driver model is not updated. The proposed strategy allows the AV to behave cautiously when there is uncertainty in the driver model and adapt its behavior according to the estimate of the model of the interacting driver.
	
	\begin{table}
	\centering
	\caption{Collision and lane change rates under various strategies. }\label{tab:MCsim}
	\begin{tabularx}{0.8\columnwidth}{@{}  Xcc @{}}
	    \toprule
		    Strategy     & Collision rate (\%) & Lane change rate (\%) \ \\ \midrule
		    Nominal  & 21          & 78                     \\ 
            Adaptive & 2             & 93                   \\ 
            Robust   & 1             & 75                    \\ 
			\bottomrule
	\end{tabularx}
\end{table}

Initializing from the same condition as mentioned above, multiple simulation runs were performed for the traffic scenario under each decision making strategy. The collision and lane changing rates under the three strategies are shown in Table \ref{tab:MCsim}. Due to space constraints, the discussions are omitted. 

	
	
	\section{Conclusions}
	\label{sec:conclusions}

	An adaptive robust decision making strategy has been proposed for the autonomous vehicles sharing the road with human drivers in a multi-agent traffic scenario. The interactions between vehicles are modeled using a level-k game-theoretic framework. The proposed robust control strategy accounts for the uncertainties in the vehicle dynamic model and the driver model estimation. The autonomous vehicle estimates the driver model of the other agents at each time step and is shown to use it to adapt its behavior through numerical simulations of a lane changing maneuver.


	\section*{Acknowledgement}

	We would like to express our appreciation to Mr. Nan Li, Dr. Ilya Kolmanovsky and Dr. Anouck Girard who provided expertise that greatly assisted the research.


	\bibliographystyle{ieeetr}
	\bibliography{Reference}

\begin{thebibliography}{10}

\bibitem{Okuda2014}
R.~Okuda, Y.~Kajiwara, and K.~Terashima, ``A survey of technical trend of adas
  and autonomous driving,'' in {\em Technical Papers of 2014 International
  Symposium on VLSI Design, Automation and Test}, pp.~1--4, IEEE, 2014.

\bibitem{Lazar2018}
D.~A. Lazar, R.~Pedarsani, K.~Chandrasekher, and D.~Sadigh, ``Maximizing road
  capacity using cars that influence people,'' in {\em 2018 IEEE Conference on
  Decision and Control (CDC)}, pp.~1801--1808, IEEE, 2018.

\bibitem{Griffiths2015}
T.~L. Griffiths, F.~Lieder, and N.~D. Goodman, ``Rational use of cognitive
  resources: Levels of analysis between the computational and the
  algorithmic,'' {\em Topics in cognitive science}, vol.~7, no.~2,
  pp.~217--229, 2015.

\bibitem{Althoff2009}
M.~Althoff, O.~Stursberg, and M.~Buss, ``Model-based probabilistic collision
  detection in autonomous driving,'' {\em IEEE Transactions on Intelligent
  Transportation Systems}, vol.~10, no.~2, pp.~299--310, 2009.

\bibitem{Stahl1993}
D.~O. Stahl, ``Evolution of smartn players,'' {\em Games and Economic
  Behavior}, vol.~5, no.~4, pp.~604--617, 1993.

\bibitem{Li2017}
N.~Li, D.~W. Oyler, M.~Zhang, Y.~Yildiz, I.~Kolmanovsky, and A.~R. Girard,
  ``Game theoretic modeling of driver and vehicle interactions for verification
  and validation of autonomous vehicle control systems,'' {\em IEEE
  Transactions on control systems technology}, vol.~26, no.~5, pp.~1782--1797,
  2017.

\bibitem{Li2018}
N.~Li, I.~Kolmanovsky, A.~Girard, and Y.~Yildiz, ``Game theoretic modeling of
  vehicle interactions at unsignalized intersections and application to
  autonomous vehicle control,'' in {\em 2018 Annual American Control Conference
  (ACC)}, pp.~3215--3220, IEEE, 2018.

\bibitem{Tian2018}
R.~Tian, S.~Li, N.~Li, I.~Kolmanovsky, A.~Girard, and Y.~Yildiz, ``Adaptive
  game-theoretic decision making for autonomous vehicle control at
  roundabouts,'' in {\em 2018 IEEE Conference on Decision and Control (CDC)},
  pp.~321--326, IEEE, 2018.

\bibitem{Scokaert1998}
P.~O. Scokaert and D.~Mayne, ``Min-max feedback model predictive control for
  constrained linear systems,'' {\em IEEE Transactions on Automatic control},
  vol.~43, no.~8, pp.~1136--1142, 1998.

\bibitem{Mayne2005}
D.~Q. Mayne, M.~M. Seron, and S.~Rakovi{\'c}, ``Robust model predictive control
  of constrained linear systems with bounded disturbances,'' {\em Automatica},
  vol.~41, no.~2, pp.~219--224, 2005.

\bibitem{Richards2003}
A.~Richards and J.~P. How, ``Model predictive control of vehicle maneuvers with
  guaranteed completion time and robust feasibility,'' in {\em American Control
  Conference, 2003. Proceedings of the 2003}, vol.~5, pp.~4034--4040 vol.5,
  June 2003.

\bibitem{Sankar2019}
G.~S. {Sankar}, R.~C. {Shekhar}, C.~{Manzie}, T.~{Sano}, and H.~{Nakada},
  ``Fast calibration of a robust model predictive controller for diesel engine
  airpath,'' {\em IEEE Transactions on Control Systems Technology}, pp.~1--15,
  2019.

\bibitem{Sankar2019a}
G.~S. Sankar, R.~C. Shekhar, C.~Manzie, T.~Sano, and H.~Nakada, ``Model
  predictive controller with average emissions constraints for diesel
  airpath,'' {\em Control Engineering Practice}, vol.~90, pp.~182 -- 189, 2019.

\bibitem{Jin2019}
G.~I. Jin, S.~Bastian, M.~M. Richard, and A.~Matthias, ``Risk-aware motion
  planning for automated vehicle among human-driven cars,'' in {\em 2019
  American Control Conference (ACC)}, pp.~--, IEEE, 2019.

\bibitem{Claussmann2015}
L.~Claussmann, A.~Carvalho, and G.~Schildbach, ``A path planner for autonomous
  driving on highways using a human mimicry approach with binary decision
  diagrams,'' in {\em 2015 European Control Conference (ECC)}, pp.~2976--2982,
  IEEE, 2015.

\bibitem{Brechtel2014}
S.~Brechtel, T.~Gindele, and R.~Dillmann, ``Probabilistic decision-making under
  uncertainty for autonomous driving using continuous pomdps,'' in {\em 17th
  International IEEE Conference on Intelligent Transportation Systems (ITSC)},
  pp.~392--399, IEEE, 2014.

\bibitem{Sadigh2016}
D.~Sadigh, S.~Sastry, S.~A. Seshia, and A.~D. Dragan, ``Planning for autonomous
  cars that leverage effects on human actions.,'' in {\em Robotics: Science and
  Systems}, vol.~2, Ann Arbor, MI, USA, 2016.

\bibitem{Kong2015}
J.~Kong, M.~Pfeiffer, G.~Schildbach, and F.~Borrelli, ``Kinematic and dynamic
  vehicle models for autonomous driving control design,'' in {\em 2015 IEEE
  Intelligent Vehicles Symposium (IV)}, pp.~1094--1099, IEEE, 2015.

\bibitem{Costa-Gomes2006}
M.~A. Costa-Gomes and V.~P. Crawford, ``Cognition and behavior in two-person
  guessing games: An experimental study,'' {\em American Economic Review},
  vol.~96, no.~5, pp.~1737--1768, 2006.

\bibitem{Costa-Gomes2009}
M.~A. Costa-Gomes, N.~Iriberri, and V.~P. Crawford, ``Comparing models of
  strategic thinking in van huyck, battalio, and beil's coordination games,''
  {\em Journal of the European Economic Association}, vol.~7, no.~2/3,
  pp.~365--376, 2009.

\end{thebibliography}

\end{document}